\documentclass{aa}
\usepackage[varg]{txfonts}
\usepackage{natbib}
\usepackage[export]{adjustbox}
\usepackage{wasysym}
\usepackage{amsmath}
\usepackage{nccmath}
\usepackage{rotating}
\usepackage{capt-of}
\usepackage{float}
\bibpunct{(}{)}{;}{a}{}{,} 

\begin{document}

\title{Dust modeling of the combined ALMA and SPHERE datasets of HD163296}

\subtitle{Is HD163296 really a Meeus group II disk?}

\author{G. A. Muro-Arena \inst{1} \and 
C. Dominik \inst{1} \and 
L. B. F. M. Waters \inst{2,1} \and 
M. Min \inst{2,1} \and 
L. Klarmann \inst{1} \and 
C. Ginski \inst{3,1} \and 
A. Isella \inst{4} \and 
M.\,Benisty \inst{5,6} \and 
A.\,Pohl \inst{7} \and 
A. Garufi \inst{8} \and 
J. Hagelberg \inst{6} \and 
M. Langlois \inst{9,10} \and
F. Menard \inst{6} \and 
C. Pinte \inst{6} \and 
E. Sezestre \inst{6} \and
G.\,van\,der\,Plas \inst{11,12} \and 
M.\,Villenave \inst{6} \and
A. Delboulb\'e \inst{6} \and 
Y. Magnard \inst{6} \and
O. M\"oller-Nilsson \inst{7} \and 
J. Pragt \inst{13} \and 
P. Rabou \inst {6} \and 
R.\,Roelfsema \inst{13} 
}

\institute{Anton Pannekoek Institute for Astronomy, University of Amsterdam, Science Park 904, 1098 XH Amsterdam, The Netherlands \and 
SRON Netherlands Institute for Space Research, Sorbonnelaan 2, 3584 CA Utrecht, The Netherlands \and 
Leiden Observatory, Leiden University, PO Box 9513, 2300 RA Leiden, The Netherlands \and 
Department of Physics and Astronomy, Rice University, 6100 Main Street, Houston, TX 77005, USA \and 
Unidad Mixta Internacional Franco-Chilena de Astronomía, CNRS/INSU UMI 3386 and Departamento de Astronomía, Universidad de Chile, Casilla 36-D, Santiago, Chile \and 
Univ. Grenoble Alpes, CNRS, IPAG, F-38000 Grenoble, France \and 
Max Planck Institute for Astronomy, Königstuhl 17, 69117 Heidelberg, Germany \and 
Universidad Autónoma de Madrid, Dpto. Física Teórica, Módulo 15, Facultad de ciencia, Campus de Cantoblanco, E-28049, Madrid, Spain \and 
CRAL, UMR 5574, CNRS, Université Lyon 1, 9 avenue Charles André, 69561 Saint Genis Laval Cedex, France \and		
Aix Marseille Université, CNRS, LAM (Laboratoire d’Astrophysique de Marseille) UMR 7326, 13388, Marseille, France \and	
Departamento de Astronomia, Universidad de Chile, Casilla 36-D, Saltiago, Chile \and  
Millenium Nucleus Protoplanetary Disks in ALMA Early Science, Universidad de Chile, Casilla 36-D, Santiago, Chile \and  
NOVA Optical Infrared Instrumentation Group, Oude Hoogeveensedijk 4, 7991 PD Dwingeloo, The Netherlands
}

\date{Received / Accepted }

\abstract{Multi-wavelength observations are indispensable in studying disk geometry and dust evolution processes in protoplanetary disks. }
{We aimed to construct a 3-dimensional model of HD\,163296 capable of reproducing simultaneously new observations of the disk surface in scattered light with the SPHERE instrument and thermal emission continuum observations of the disk midplane with ALMA. We want to determine why the SED of HD\,163296 is intermediary between the otherwise well-separated group I and group II Herbig stars.}
{The disk was modelled using the Monte Carlo radiative transfer code \textit{MCMax3D}. The radial dust surface density profile was modelled after the ALMA observations, while the polarized scattered light observations were used to constrain the inclination of the inner disk component and turbulence and grain growth in the outer disk.}
{While three rings are observed in the disk midplane in millimeter thermal emission at $\sim$80, 124 and 200 AU, only the innermost of these is observed in polarized scattered light, indicating a lack of small dust grains on the surface of the outer disk. We provide two models capable of explaining this difference. The first model uses increased settling in the outer disk as a mechanism to bring the small dust grains on the surface of the disk closer to the midplane, and into the shadow cast by the first ring. The second model uses depletion of the smallest dust grains in the outer disk as a mechanism for decreasing the optical depth at optical and NIR wavelengths. In the region outside the fragmentation-dominated regime, such depletion is expected from state-of-the-art dust evolution models. We studied the effect of creating an artificial inner cavity in our models, and conclude that HD\,163296 might be a precursor to typical group I sources.}
{}


\maketitle

\section{Introduction}

Protoplanetary disks are the sites of planet formation, of which the disks around Herbig Ae/Be are the best studied examples due to their brightness and proximity to Earth. The structure and morphology of protoplanetary disks around Herbig Ae/Be stars is as of now not well understood, though different models have been proposed over the years to link their global geometry to the characteristics of their spectral energy distributions (SEDs). The classification system initially proposed by \citet{2001A&A...365..476M} first divided Herbig Ae/Be SEDs into two groups based on the far infrared excess of these sources: those whose SEDs can be fitted by a single powerlaw component (group II), and those that require an additional broad component peaking in the mid-to-far infrared (group I). An additional subdivision was proposed for group I based on the presence (group Ia)or absence (group Ib) of the 10 $\mu m$ silicate feature, whereas it was observed that this silicate feature was present in all group II sources.

The difference in the SEDs between group I and group II sources was for a long time attributed to a flaring versus flat geometry \citep{2004A&A...417..159D}, the larger infrared excess of group I sources being explained by the absorbed and reprocessed light in the flaring outer regions of these disks. \citet{2013A&A...555A..64M} on the other hand point at the inner wall of a flaring outer disk, directly illuminated by the central star due to the presence of a large gap or inner cavity in the disk, as the origin of the cold component in group I SEDs, with the near-infrared excess attributed to an optically thin and hot component at the interior of this cavity. We now believe group II sources to be either flat of very compact objects, due to the fact that they can rarely be resolved in scattered light \citep{2017A&A...603A..21G}.

Though classically classified as a group II source \citep{2001A&A...365..476M}, HD\,163296 is sometimes referred to as an intermediate-type source in the literature (e.g. \citealt{2003A&A...398..607D}), with a far infrared excess greater than typical group II sources yet lower than group I sources. If the view we hold of group II sources as compact or flat, self-shadowed disks is correct, then direct observations with the \textit{Hubble Space Telescope} (\textit{HST}) \citep{2000ApJ...544..895G, 2008ApJ...682..548W} point towards a possible misclassification of the disk around HD\,163296. Dust in the disk has been detected in scattered light by both studies out to radial distances of at least $\sim$450 AU, direct evidence that the disk is not compact and the outermost regions are flaring and directly illuminated by the central star. Recently in a taxonomical study of a sample of 17 group I and group II sources by \citet{2017A&A...603A..21G} it has been proposed that HD\,163296 might be a precursor to typical group I disks. 

HD\,163296 is a $\sim$5.1$^{0.3}_{0.8}$\,Myr old \citep{2013MNRAS.429.1001A} Herbig Ae star located at $122^{+17}_{-13}$\,pc from the Sun \citep{1997A&A...324L..33V}. Its mass is $2.23^{0.22}_{0.07}$\,M$_{\odot}$ \citep{2013MNRAS.429.1001A}, and it has a luminosity of $\sim$33.1$^{6.7}_{2.3}$\, L$_{\odot}$ \citep{2013MNRAS.429.1001A}. With an effective temperature of $\sim$9200$\pm300$\,K \citep{2012MNRAS.422.2072F}, it is classified as a spectral type A1Ve star \citep{1998A&A...330..145V}. The massive disk surrounding it was first resolved interferometrically twenty years ago by \citet{1997ApJ...490..792M} using the Owens Valley Radio Observatory (OVRO) millimiter-wave array, with sizes derived for both the continuum and gas line emission at $1.3$\,mm of $\sim$100 and 300 AU, respectively. Recent, more sensitive observations with ALMA have since shown the gaseous component of the disk extends to distances of at least $500$\,AU \citep{2016PhRvL.117y1101I, 2013A&A...557A.133D}, while the dust continuum at millimeter wavelengths is only detected out to a radial distance of $\sim240$\,AU from the central star. The continuum images reveal further structure in the dust: a bright inner disk component is observed within the inner $\sim$0.5 arcsec (60 AU) of the image, with two bright rings at $\sim$0.66 arcsec (80 AU) and $\sim$1.0 arcsec (122 AU) and a third much fainter ring at $\sim1.6$ arcsec (195 AU) \citep{2016PhRvL.117y1101I}. 

In addition to the \textit{HST} observations, ground-based observations in polarized scattered light in the J-Band with the Gemini Planet Imager \citep{2017ApJ...838...20M} and in H- and Ks-band with VLT/NACO \citep{2014A&A...568A..40G} have also shown emission in the near infrared out to distances of $\sim$80 AU around HD\,163296, much more compact than the scattered light seen in \textit{HST} images yet providing additional evidence that the disk around HD\,163296 is not flat and self-shadowed, or at least not at all radii. This is also smaller than the disk observed in the millimeter (mm) continuum, which poses an interesting question as to the nature of this and possibly other so-called intermediate sources which might not fit easily with current interpretations of the group classification scheme. The drastically different nature of the observed disk structure between NIR and mm-wavelengths observations shows that neither stands on its own as a complete, comprehensive picture of this disk. With this in consideration, we have attempted to produce a single, physically-consistent 3-dimensional model able of simultaneously reproducing the dust observations in both NIR and mm-wavelenghts. 

In this paper we present new polarized scattered light observations of HD\,163296 obtained with SPHERE/IRDIS and describe the data reduction in Sect. \ref{Section2}. Results of these observations are presented in Sect. \ref{Section3}. In Sect. \ref{Section4} we detail the procedure followed to model the surface density of the disk based on the ALMA observations \citep{2016PhRvL.117y1101I}, and propose two different models capable of explaining some of the features seen in the SPHERE/IRDIS data. The results are discussed in Sect. \ref{Section5}, particularly in the context of disk geometry and its role in the Meeus group classification scheme. Finally, we present our conclusions and summarize our findings in Sect. \ref{Section6}.

\section{Observations}\label{Section2}

\subsection{Data reduction}

HD\,163296 was observed as part of the SPHERE \citep{Beuzit2008} guaranteed time program with the infrared dual band imager and spectrograph (IRDIS, \citealt{2008SPIE.7014E..3LD}) on May 26th 2016. Observations were carried out in dual-polarization imaging mode (DPI, \citealt{2014SPIE.9147E..1RL}) in J and H-band. The sky was clear during the observations, but seeing conditions were variable with an average seeing of 0.9\,arcsec in the optical and seeing spikes up to 1.2\,arcsec, especially during the H-band sequence. In DPI mode IRDIS uses polarizers inserted into the beam to measure two orthogonal polarization directions simultaneously (ordinary and extraordinary beam). A half-wave plate (HWP) is used to rotate the polarisation of the incoming light beam such that the two linear Stokes vector components Q and U can be meassured. The HWP is also used to correct in first order for instrumental polarization by rotating in steps of 22.5$^\circ$. Such HWP rotation sequence allows to measure the four linear polarization components per polarization cycle: Q$^+$, Q$^-$, U$^+$ and U$^-$. The HWP orientation is choosen such that between Q$^+$ and Q$^-$ only the sign of the astrophysical signal of the Stokes Q component changes, while the instrumental polarization downstream from the HWP remains unchanged. Thus by subtracting Q$^-$ from Q$^+$ instrumental polarization is cancelled out while the astrophysical signal is preserved. For J-band we recorded a total of 15 polarization cycles with an individual exposure time (DIT) of 16\,s and 5 exposures per HWP position (NDIT). This amounted to a total integration time of 80\,min. In H-band we recorded 3 polarization cycles, one cycle with a DIT of 8s and an NDIT of 16 and two cycles with a DIT of 16s and and NDIT of 8. The change in exposure time was performed to adjust to the variable Seeing conditions. This amounted to a total integration time of 25.6\,min in H-band.\\

\begin{figure*}[ht]
  \centering
    \includegraphics[width=180mm]{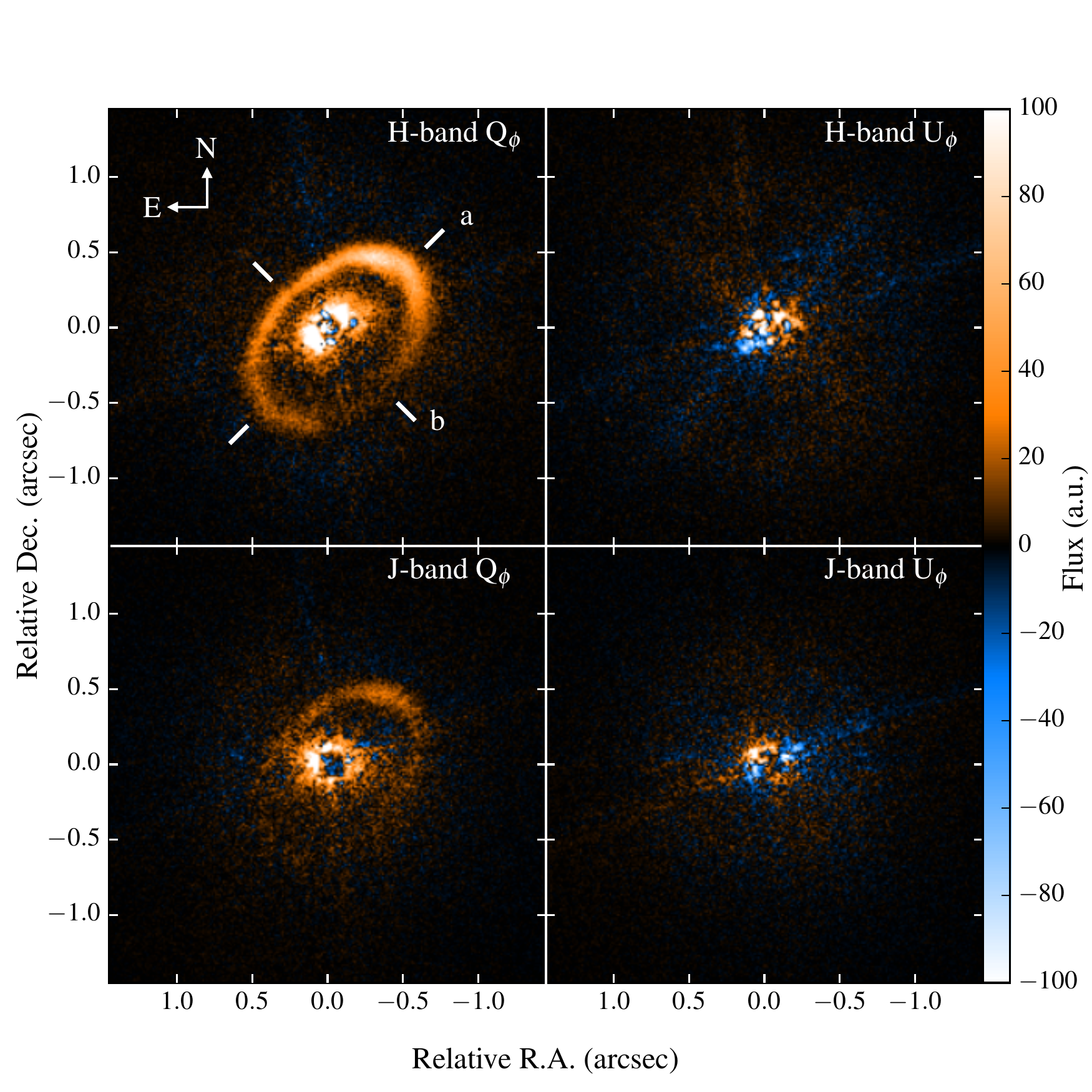}
  \caption{H-band (top) and J-band (bottom) DPI observations of HD\,163296 with SPHERE/IRDIS. Left column shows the \textit{Q$_{\phi}$} images, with the right column showing \textit{U$_{\phi}$}. A single asymmetrical ring is obserbed in scattered light at $\sim$0.6 arcsec from the center of the disk in both Q$_{\phi}$ images, slightly offset from center in the South-West direction. An additional bright inner component shows up in both bands, the center of which is obscured by the coronograph. Some residual signal shows up in both U$_{\phi}$ images, mostly constrained to the inner 0.2-0.3 arcsec. The white dashes indicate the position of the major and minor axes, labeled a and b respectively.}\label{SPHERE}
\end{figure*}

The data reduction was performed following largely the approach layed out in \cite{2016A&A...595A.112G}. We give here a short summary and highlight differences to the mentioned procedure. After dark, bad pixel and flat field correction, we performed a single difference step by subtracting ordinary and extraordinary beams from each other, thereby creating the previously mentioned Q$^+$, Q$^-$, U$^+$ and U$^-$ polarization components. We then performed the so called double difference step (\citealt{2001ApJ...553L.189K}, \citealt{2004IAUS..221..307A}) by subtracting Q$^-$ from Q$^+$ (and analog for U) to generate the Stokes Q and U components for each polarization cycle. Since conditions were variable and thus the tip-tilt correction of the AO system was not always stable, we ensured a good alignment of the frames before the double difference step. For this purpose we utilized several bright background point sources to which we fitted Moffat functions to measure their positions between frames. After the double difference step we combined the Q and U frames for all polarimetric cycles. We then performed an instrumental polarization correction as described by \cite{2011A&A...531A.102C} to subtract the residual instrumental polarization from the final Stokes Q and U frames. We assumed for this purpose that the stellar light is completely unpolarized. We thus measured the stellar signal at the bright adaptive optics correction cut-off radius in Stokes Q, U and the intensity image and then subtracted a scaled version of the intensity image from the Q and U images to surpress remaining signal which should be instrumental in nature. After this final step was performed we computed the polarized intensity (PI) image from the Stokes Q and U components:

\begin{ceqn}
 \begin{align}
    PI = \sqrt{Q^2 + U^2}.
 \end{align}
\end{ceqn}

Since the disk around HD\,163296 is not strongly inclined, it is reasonable to assume that the polarized light that we receive from the system is overwhelmingly created in single scattering events \citep{2015A&A...582L...7C}. We can thus assume that the angle of linear polarization shows an azimuthal alignment with respect to the central star. It is then convenient to use the polar Stokes components Q$_\phi$ and U$_\phi$ as defined by \cite{2006A&A...452..657S}:

\begin{ceqn}
 \begin{align}
	Q_\phi &= Q \times \cos{2\phi} + U \times \sin{2\phi}, \\		
	U_\phi &= Q \times \sin{2\phi} - U \times \cos{2\phi},
 \end{align}
\end{ceqn}

\noindent wherein $\phi$ is the azimuth with respect to the center of the star. Q$_\phi$ contains all the azimuthal polarized flux as positive signal, whereas U$_\phi$ contains all polarized flux with a 45$^\circ$ offset from azimuthal. Thus U$_\phi$ is in our case expected to contain very little or no signal and can be used as a convenient noise estimator. We show the final reduced Q$_\phi$ and U$_\phi$ images for both bands in Fig~\ref{SPHERE}.

\subsection{Observational Results}\label{Section3}

The Q$_\phi$ images of HD\,163296 in both J and H-band in Fig~\ref{SPHERE} show the same ring structure at approximately 0.6 arcsec from the center of the image, with only the North half of the ring clearly visible in the J-band image due to a lower contrast. The ring appears to have an offset from the center of the image in the South-West direction, likely a combined product of the inclination of the disk and the scattering surface being located at a certain height from the midplane, indicative of a geometrically-thick disk \citep{2016A&A...595A.112G, 2016A&A...595A.114D}. It can be inferred from the direction of the offset that the North-East side of the disk corresponds to its front side.

The polarized flux density in the ring appears asymmetric in both J and H bands. First, there is the usual asymmetry in the direction of the minor axis, likely to be caused by the different scattering angles for the front part of the disk (North-East, brighter because of forward scattering) and back part (South-West, darker), in combination with the scattering phase function. In addition there is also an asymmetry along the major axis in both bands, with the brightest side of the ring corresponding to the North-West side and roughly but not exactly coincidental with the major axis. The asymmetry along the major axis cannot be explained by geometric effects and the scattering phase function. Without other factors at play, and assuming the dust is distributed homogeneously in the ring, the measured flux density is expected to be symmetric along the major axis (i.e. mirror-symmetric w.r.t. the minor axis). The observed asymmetry is thus probably the result of either an uneven dust distribution in the surface of the disk, an azimuthally-variable scale height, or the product of a shadow cast on the surface on one side of the ring by an inner disk component. 

Elliptical annuli were fitted to both J and H-band Q$_{\phi}$ images to obtain the semi-major axis $a$, position angle PA and offset of the ring ($\Delta x_{0}$, $\Delta y_{0}$). The fitting procedure consisted of a Monte Carlo routine that generated $10^{6}$ elliptical annuli with all four parameters randomized within restricted ranges. The width of each annulus was set to 4 pixels, and the inclination was set at $i=\,46^{\circ}$ for both H- and J-band images. The total flux was averaged over each aperture, and the best fit parameters were obtained from the elliptical ring aperture which maximizes the averaged flux. We chose to fix the inclination at this given value, corresponding to the inclination measured from the ALMA image, as most attempts to measure the inclination from the SPHERE images yielded inclinations larger than $i=\,52^{\circ}$ for the J-band image and similarly too-large values for the H-band. This resulted in apertures which did not properly fit the far side of the ring, most likely due to the variable width and brightness of the ring in both images, with the far side of the ring appearing both wider and fainter than the near side. 

The errors for each parameter were obtained from the H-band image, as it has a higher signal-to-noise ratio. The noise was measured from the U$_{\phi}$ image as the standard deviation $\sigma_{U_{\phi}}$ of the flux inside the best-fitting aperture. The uncertainty the noise introduces in the flux measured from the Q$_{\phi}$ image was calculated as $\Delta F= \sigma_{U_{\phi}}/\sqrt N$, where $N$ is the number of pixels in the aperture. We then select all those apertures from the initial 10$^{6}$ whose average flux is larger than $F_{max} - \Delta F$, and obtain the errors in the four fitted parameters as the standard deviation of the parameters for this sub-set of apertures. 

A semi-major axis of $a=$ 0.63$\pm$0.045 arcsec was measured in the H-band, which corresponds to a radius of 77$\pm$5.5 AU (122 AU per arcsecond, assuming a distance of 122 pc), along with a position angle of PA= $134.8^{\circ}\pm11^{\circ}$. The ring observed in the SPHERE images thus matches the size and orientation of the first ring observed in the Band 6 ALMA continuum observations, which has a radius of $\sim$0.66 arcsec or 80 AU and a position angle of PA= $134^{\circ}$. The large error measured for the position angle is again likely caused by the uneven width and brightness around the ring, which introduces uncertainty in our fitting procedure by allowing a broad range of position angles to yield similar fluxes inside the apertures. An offset of 105$\pm$45 mas was measured in the H-band in the South-West direction (PA=\,$215^{\circ}\pm22^{\circ}$), in good agreement with the offset measured by \citet{2017ApJ...838...20M} and corresponding to a height of the scattering surface of 12.7$\pm$5.5 AU above the disk midplane. A similar offset of 92 mas was measured in the J-band image, which gives a scattering surface height of 11.2 AU above the midplane. The semi-major axis measured in the J-band was $a=$\,0.61 arcsec or 74 AU. The position angle measured in this band is PA=\,$137.5^{\circ}$.

No flux with an SNR above 3 is measured in either J or H-band beyond $\sim$100 AU, in contrast to the second and third rings detected in the ALMA continuum observations out to distances of $\sim$240 AU, but confirming the non-detections at larger radii by \citet{2017ApJ...838...20M,2014A&A...568A..40G}, who also observe a single ring with a similar asymmetry to it. Interestingly, no emission is observed out to larger radii of $\sim$500 AU, whereas \textit{HST} did detect scattered light in the disk out to these radii \citep{2000ApJ...544..895G, 2008ApJ...682..548W}. This is likely due both to the higher sensitivity of \textit{HST} to faint, extended features and the fact that it observes total intensity as opposed to the polarized light of the SPHERE images. A bright inner disk component can be seen in the H-band Q$_{\phi}$ image and less prominently in the J-band Q$_{\phi}$ image, in both cases partially obscured by the coronograph. 

\begin{figure}[t]
  \centering
    \includegraphics[width=88mm,left]{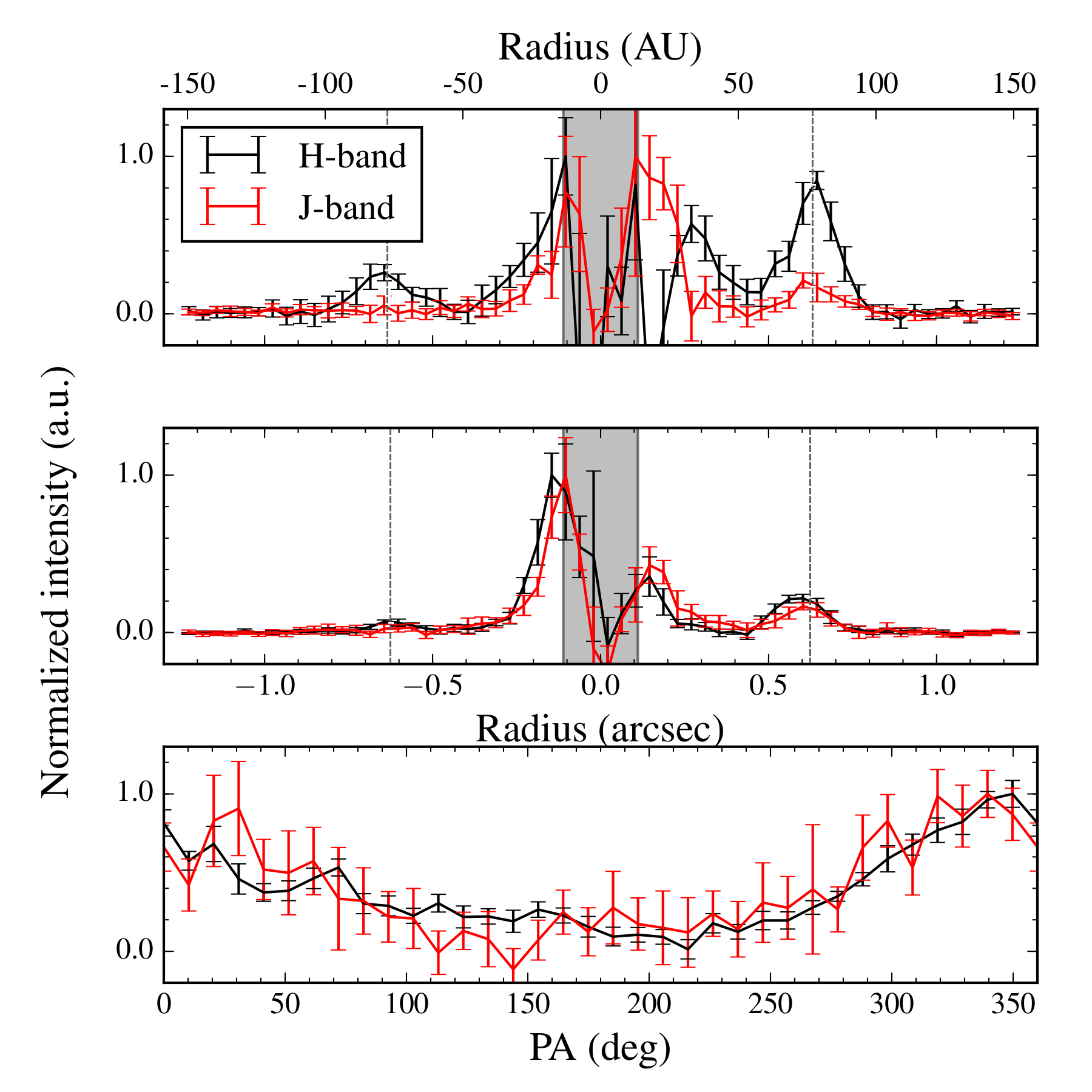}
  \caption{Top: radial polarized intensity profile through the center of the ring along the major axis for H-band (black) and J-band (red) Q$_{\phi}$ images. Middle: radial polarized intensity profile through the star in the direction of the major axis for both bands. Positive radii correspond to the North-West direction, and negative radii to the South-East direction. The vertical dashed lines mark the approximate location of the ring at 77 AU. The grey region at the center of the figure corresponds to the location of the coronograph. Bottom: azimuthal intensity profile of the ring in H-band (black) and J-band (red) at a distance of 77 and 74 AU from the center of the disk, respectively, centered at the center of each ring. The position angle PA is measured East-from-North. All profiles are normalized to the maximum intensity.}\label{SPH_profile}
\end{figure}

The top panel of Fig~\ref{SPH_profile} shows the radial profiles of both Q$_{\phi}$ images along the ring's major axis. Each data point corresponds to the flux density averaged over all the pixels contained in an aperture with a radius equal to the image resolution, for each band, and then normalized to the maximum value of the profile. The error bars correspond to the standard deviation of the flux density at each aperture over the square root of the number of point spread functions (PSFs) contained in it. The panel shows the brightness asymmetry in the ring along this axis in both photometric bands, with positive radii corresponding to the North-West direction and negative radii to South-East. The H-band profile shows the North-West side of the ring is brighter than the South-East side by about a factor 3. The South-East side of the ring is not visible in the J-band image, but the position of the North-West side of the ring matches that of the H-band image. The ring width is well-resolved in both bands, with the FWHM of the ring approximately four times the image resolution at either end of the major axis. The profile of the inner disk is marred by the small circular (instrumental) artifacts that can be seen close to the center of the image in both bands in Fig~\ref{SPHERE}: a single one in the J-band, and three in the H-band in a triangular formation around the coronograph, two of them along the ring major axis. The middle panel of Fig~\ref{SPH_profile} shows the radial profile through the position of the star in the direction of the ring major axis, which provides a better view of the profile of the inner disk. In both images it can be observed that the inner disk appears brighter on the South-East side of the coronograph.

The bottom panel of Fig~\ref{SPH_profile} shows the azimuthal profile of each Q$_{\phi}$ image at the location of the ring as a function of the position angle, measured East-from-North. The data again correspond to the pixel-averaged flux density inside apertures with the same radii as described above, normalized to the maximum of the profile. Error bars correspond to the standard deviation of the flux density at each aperture over the square root of the number of PSFs contained in the apertures. It is observed that the maxima both occur at a position angle of $\sim$350$^{\circ}$, about 35$^{\circ}$ from the semi-major axis towards the front side of the disk. The profiles are in good agreement with each other within the errorbars except at position angles between 80$^{\circ}$ and 180$^{\circ}$, where the ring vanishes in the noise in the second quadrant of the J-band image whereas the ring still apears clearly visible at this location in the H-band image.

\section{Modelling the dusty disk around HD\,163296}\label{Section4}

The observations with the SPHERE instrument differ greatly from the disk observed in ALMA band 6 images \citep{2016PhRvL.117y1101I}. In contrast to the single ring at $\sim$77 AU observed in the J and H-bands, the ALMA continuum and line images show a disk which extends at least out to around 500 AU in gas and 240 AU in large dust grains \citep{2013A&A...557A.133D,2016PhRvL.117y1101I}, and presents both a large inner disk extending out to $\sim$50 AU and three outer rings. In this section we present the results of two models capable of explaining the differences observed between the midplane thermal emission observations and the scattered light observations of the disk surface. We do this in the context of global disk geometry and the how it ties into the Meeus group classification scheme.

\subsection{The initial model}\label{model}

The modelling of the disk was done using the three-dimensional Monte Carlo radiative transfer code \textit{MCMax3D} \citep{2009A&A...497..155M}. The modelling of two independent zones is necessary to fit the disk SED; a first zone composed of an unresolved inner disk (UID) to account for the NIR excess in the SED, and a second zone to model the structure resolved by the polarized scattered light and ALMA continuum observations. The UID has been spatially resolved by NIR interferometric data \citep{2010A&A...511A..74B, 2017A&A...599A..85L} and modeled independently by several studies \citep{2017A&A...599A..85L, 2010A&A...519A..26R, 2008ApJ...689..513T}. The focus of this work is on the outer disk resolved by the SPHERE and ALMA data, however, and the model of the UID presented here is not tested by the current observations.

The \textit{MCMax3D} code allows the user to define disk zones with different pressure scale heights, surface densities, grain size distributions, gas-to-dust ratios and turbulences. The code uses a powerlaw pressure scale height profile of the form:

\begin{ceqn}\label{H_g1}
 \begin{align}
    H_{g}(r) = H_{0} \left( \frac{r}{r_{0}} \right)^{p} ,
 \end{align}
\end{ceqn}

\noindent where $H_{0}$ is the scale height of the disk at a radius $r_{0}$ and $p$ is a positive exponent. A simple powerlaw surface density profile is defined for the UID:

\begin{ceqn}\label{sigma}
 \begin{align}
    \Sigma(r) \propto r^{-\epsilon} ,
 \end{align}
\end{ceqn}

\noindent with $\epsilon$ a positive number. The resolved disk is modeled using a powerlaw profile with an exponential taper-off:

\begin{ceqn}\label{sigma2}
 \begin{align}
    \Sigma(r) \propto r^{-\epsilon} \exp \left( - \left( \frac{r}{r_{\rm c}} \right)^{2-\gamma} \right) ,
 \end{align}
\end{ceqn}

\noindent where $r_{\rm c}$ is the taper-off radius, and $2- \gamma$ is a positive exponent, while the grain size distribution is modeled as:

\begin{ceqn}\label{na}
 \begin{align}
    n(a)\ da \sim a^{-p_{a}}\ da ,
 \end{align}
\end{ceqn}

\noindent for a particle size $a$ between a minimum and maximum particle sizes $a_{\rm min}$ and $a_{\rm max}$, respectively, and $p_{a}$ a positive exponent. 

A two-zone model from the DIANA project database \citep{2016A&A...586A.103W}, obtained from fitting the spectral energy distribution with a genetic fitting algorithm, was used as a starting point for our modeling (P. Woitke, private communication, http://www.univie.ac.at/diana/index.php/user/login). The parameters characterizing this model can be found in Table \ref{Tab1}. A gas-to-dust ratio of a 100 is assumed for the whole disk, while the turbulence is characterized by $\alpha \sim 2.4\times 10^{-3}$, which is in agreement with the upper limit of $\alpha = 3\times 10^{-3}$ reported by \citet{2017arXiv170604504F}. Both zones are assumed to have the same inclination and position angle. For this model we have used the values measured from the first ring in the ALMA image, PA=$132^{\circ}$ and $i=46^{\circ}$.

\begin{table}
 \caption{Initial \textit{MCMax3D} model parameters}\label{Tab1}
 \centering
 \begin{tabular}{l@{\hskip 0.15in}c@{\hskip 0.15in}c}\hline
     & Zone 1 & Zone 2\\\hline\hline
 R$_{\rm in}$ (AU) & $4.04\times 10^{-1}$& 1.70\\\hline
 R$_{\rm out}$ (AU) & 1.07 & 240\\\hline
 M$_{\rm d}$ (M$_{\odot}$)          & $6.6\times 10^{-9}$   & $4.23\times 10^{-4}$\\\hline
 $\epsilon$ & $7.35\times 10^{-1}$& 1.035\\\hline
 $\gamma$ & - & 0.585\\\hline
 r$_{\rm c}$ (AU) & - & 120\\\hline
 H$_{0}$ (AU) & $4.83\times 10^{-2}$& 5.64\\\hline		
 r$_{0}$ (AU) & 1 & 100\\\hline
 p & 1.37 & 1.15 \\\hline
 a$_{\rm min}$ ($\mu$m) & $5\times 10^{-2}$& $5\times 10^{-2}$\\\hline
 a$_{\rm max}$ ($\mu$m) & $9.75\times 10^{-1}$& $1\times 10^{4}$\\\hline
 p$_{a}$ & 3.795 & 3.795\\\hline
 g2d & 100 & 100\\\hline
 $\alpha$ & $2.4\times 10^{-3}$ & $2.4\times 10^{-3}$\\\hline
 \end{tabular}
\end{table}

\begin{figure}[t]
  \centering
    \includegraphics[width=88mm,left]{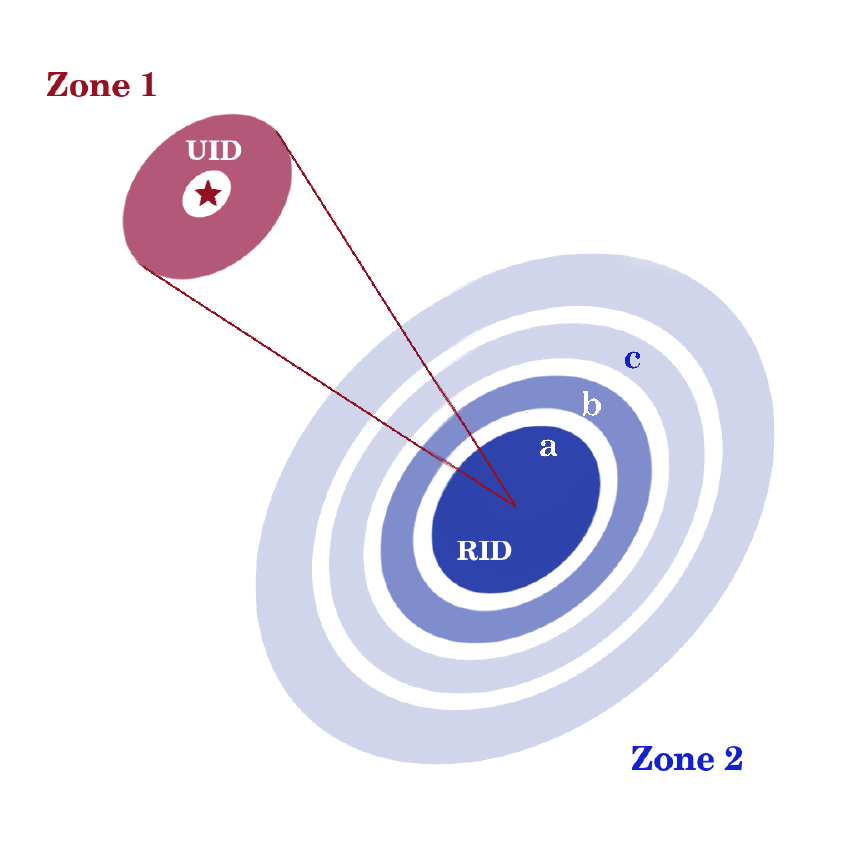}
  \caption{Graphical representation of the 2-zone \textit{MCMax3D} model. Zone 1, in red, corresponds to an unresolved inner disk component with an outer radius of 1 AU. Zone 2 corresponds to the outer disk resolved by the ALMA observations. Zone 2 has been subdivided into zones 2a, 2b and 2c, corresponding to the resolved inner disk (zone 2a), inner ring (zone 2b), and second and third rings (zone 2c). This subdivision was created for the purpose of locally altering some model parameters. For the purpose of modelling the surface density of the disk, however, Zone 2 is treated as a single region. The inner and outer radii for each zone are given in Tables \ref{Tab1} and \ref{Tab1_2}.}\label{zones}
\end{figure}

A graphical representation of the model zones is shown in Fig~\ref{zones}. The UID is shown in red (zone 1), with the resolved disk shown in shades of blue (zone 2). Zone 2 was additionally sub-divided into zones 2a, 2b and 2c. These zones account for the RID seen in the ALMA image (zone 2a), the first ring seen in both ALMA and SPHERE images (zone 2b), and the two outer rings seen only in the ALMA image (modeled jointly in zone 2c). The inner and outer radii for these zones are listed in Table \ref{Tab1_2}. Below we will discuss the different steps used to arrive at a full model of the HD\,163296 disk. Starting from the initial model, we will model the surface density in order to match the ALMA observations (Section \ref{surfdensmod}). While this will be successful, it turns out that in order to match the SPHERE observations as well, modifications are required in zone 2. These modifications will be described in Sections \ref{alpha} to \ref{misallignment}. All initial parameters for zones 2a, 2b and 2c in our models correspond to the parameters of zone 2 listed in Table \ref{Tab1}. For consistency, we shall refer to the different zones in the disk instead of disk components from now onwards when referring to the models.

\begin{table}
 \caption{Inner and outer radii of zones 2a, 2b and 2c in our model.}\label{Tab1_2}
 \centering
 \begin{tabular}{l@{\hskip 0.15in}c@{\hskip 0.15in}c@{\hskip 0.15in}c}\hline
     & Zone 2a & Zone 2b & Zone 2c\\\hline\hline
 R$_{\rm in}$ (AU) & 1.70 & 63 & 105\\\hline
 R$_{\rm out}$ (AU) & 63 & 105 & 240\\\hline
 \end{tabular}
\end{table}

\subsection{Surface density}\label{surfdensmod}

The SPHERE/IRDIS images are an example of how polarized scattered light images do not constitute the best tracer for the dust \textit{mass} in the disk. Compared to the millimeter emission seen in the ALMA image, the disk appears to be much reduced in size, with all of the structure outside the ring at 0.63 arcsec (77 AU) completely invisible under the noise level. This is likely due in part to the low percentage of the total intensity that is polarized. The emission in the RID is affected by leftover instrumental artifacts, which limit the fidelity with which we can reconstruct the surface density of the disk in this region. 

The thermal emission we observe at 1.3 mm in the ALMA continuum image delivers a more robust view of the bulk of the dust mass. It is not subject to shadowing effects and it is not affected by the instrumental artifacts caused by the coronograph and the bright point source at the center of the image, though it does depend on the temperature structure of the disk and its optical depth, and the radial extent of the disk is limited by radial drift. The disk is extremely bright at millimeter wavelengths and due to the high sensitivity of ALMA, the signal-to-noise ratio is also very large. With these considerations, we can use the spatial brightness distribution of the disk as observed in the ALMA image to model the radial surface density profile of the dust to high fidelity, limited only by the spatial resolution of the image and dependent on the parameters used for the initial model. 

This modelling of the surface density is done iteratively since the RID is optically thick even at millimeter wavelengths, which translates to non-linear variations of the surface brightness of the disk as a function of the surface density. Also, changing the surface density influences the temperature in the disk, requiring iteration. The process followed can be summarized in the following steps, assuming a perfectly azimuthally symmetric surface density distribution: 

\begin{figure*}[ht]
  \centering
    \includegraphics[width=180mm,left]{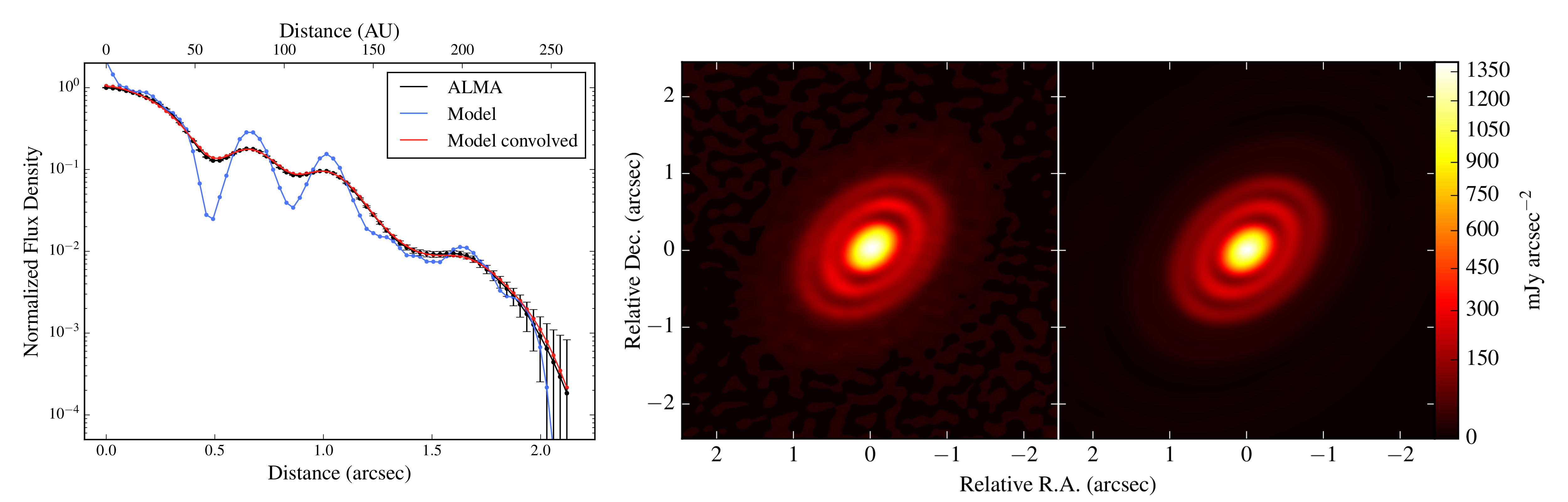}
  \caption{Left: Radial profiles of the ALMA image (black), raytraced model image (blue), and raytraced model image after convolution with a 2-dimensional Gaussian kernel with dimensions of 0.22 arcsec by 0.15 arcsec and a PA of -88$^{\circ}$ (green). Each point was obtained by integrating the flux density in elliptical annuli of varying radii and normalizing first by the aperture area, and then by the maximum of the ALMA image profile. The error bars for the ALMA profile were obtained as the rms of the image divided by the number of synthesized beams contained in each annuli. Right: Side-by-side comparison of the ALMA image (left) and the raytraced image of one of our models (right) after Gaussian blurring.}\label{ALMA_profile}
\end{figure*}

\begin{itemize}
 \item The radiative transfer model for the initial set of parameters (described in \ref{model}) is obtained with \textit{MCMax3D} for a given surface density profile, starting with the powerlaw surface density characterized by the parameters in Table \ref{Tab1}.
 \item The obtained model is raytraced to obtain an image of the disk at 1.3 mm.
 \item The raytraced image is convolved with a Gaussian kernel to simulate the ALMA reconstructed image. The kernel shape and orientation are chosen to match the characteristics of the synthesized beam of the ALMA observations: beam dimensions are 0.22 arcsec x 0.15 arcsec, with a PA of $-88^{\circ}$ measured North-to-East.
 \item The radial profile of the ALMA image is obtained by integrating the disk flux in elliptical annuli of increasing size centered on the center of the image, and normalized by the annulus area. The elliptical annuli have a PA of 132$^{\circ}$, eccentricity of 0.719 (corresponding to an inclination of 46$^{\circ}$) and a width of 5 pixels.
 \item The radial profile of the convolved model image is obtained in the same manner.
 \item The radial surface density profile used as input for the model is then scaled by the point-by-point ratio between the ALMA brightness profile and the model brightness profile. In this way mass is added to the disk at radii at which the model is too faint, and mass is substracted at radii at which the model is too bright (i.e. gaps).
 \item This scaled surface density is then used as the starting point for the next iteration of the model. 
\end{itemize}

By iterating the steps above to model the surface density of Zone 2 a radial intensity profile can be obtained that converges to that of the ALMA image profile, as seen in the left panel of Fig~\ref{ALMA_profile}, which shows an excellent agreement between the radial intensity profile of the ALMA observation and our model (using the parameters in Table \ref{Tab1}) after the iteration of the surface density. We have called this model M0. A side-by-side comparison of the ALMA observations and the convolved image of the model can be seen in the right panel of Fig~\ref{ALMA_profile}, which shows excellent overall agreement between the ALMA data and model M0. Both the data and the model appear to show an RID with a slightly lower PA than the rest of the disk, but this is an effect of the orientation of the synthesized beam and the Gaussian kernel that was used to convolve the raytraced model image. This method is only applicable to azimuthally symmetric sources, and the results are also of course model-dependent. A different grain size distribution, turbulence, chemical composition or scale height of one of the disk components, for example, will produce a different resulting surface density profile. In addition, it can only be used as far as the larger grains extend out in the disk. The ALMA observations provide us with the ability of measuring the surface density out to radii of $\sim$240 AU, as no thermal emission is detected in the continuum image above the noise level beyond that point. The model we present here is thus truncated at 240 AU, despite the observational evidence of both gas and small dust grains on the disk surface extending out to over 500 AU. 

\begin{figure*}[]
 \centering
  \includegraphics[width=128mm]{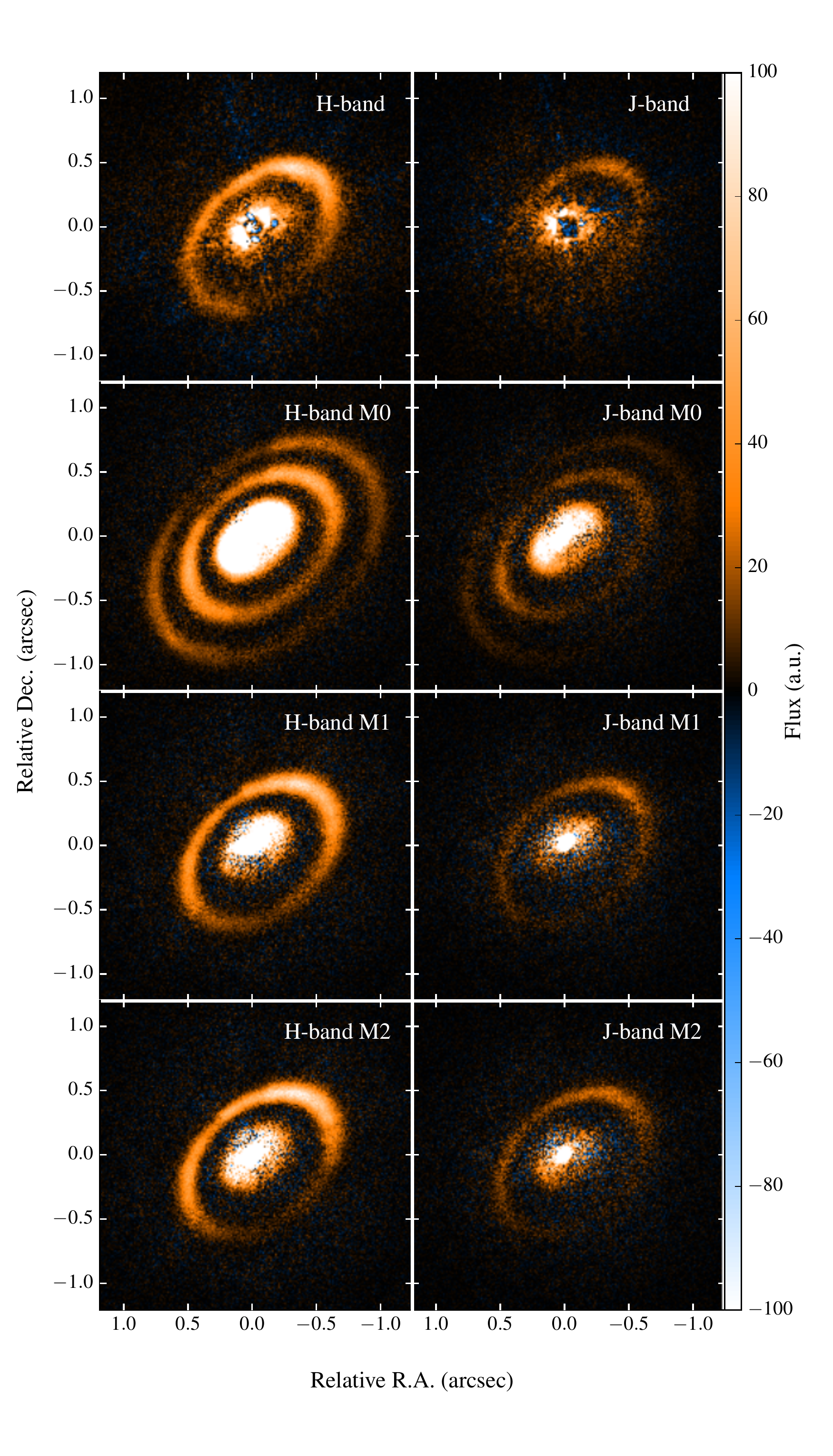}
  \caption{Side-by-side comparison of the Q$_{\phi}$ scattered light images for both H-band (left) and J-band (right) for our observations (top), model M0 (second row), model M1 (third row) and model M2 (bottom). Model M0 consists of the model characterized by the parameters in Table \ref{Tab1}. Model M1 consists of a UID inclined by 3$^{\circ}$ with respect to the outer disk, characterized by an inclination $i$=\,45$^{\circ}$ and position angle PA=\,136$^{\circ}$, along with a small grain depletion characterized by a minimum grain size a$_{\rm min}$=\,3\,$\mu$m in zone 2c. Model M2 consists of inclined UID and RID components inclined by 1.3$^{\circ}$ with respect to the rest of the disk. This inclination is characterized by $i$=\,45$^{\circ}$ and PA=\,133.2$^{\circ}$. Increased settling in the outer disk of this model is characterized by $\alpha=\, 1\times10^{-5}$ in zone 2c.}\label{Asymmetry}
\end{figure*}

The width and depth of the gaps observed in the disk structure cannot be uniquely determined from the observations, as these are limited by resolution. \citet{2016PhRvL.117y1101I} study the gap width-depth degeneracy in detail, and offers a more comprehensive view on the extent of this degeneracy, based on the assumption that the surface density of the disk can be approximated as a powerlaw with square gaps in it. The method described here offers more degrees of freedom in that the overall dust surface density of the disk is not constrained to a radial powerlaw, and in that it allows for a smooth and continuous radial surface density profile which is probably a more natural approximation to the real surface density of the disk than that of sharp gap edges. This is supported by the radial profile observed in the J and H-band images in Fig~\ref{SPH_profile}, which have an angular resolution $\sim$4 times higher than that of the ALMA image and in which no sharp edges are detected for the ring at 77 AU. 

While this model is successful at reproducing the data at milimeter wavelengths, it fails in the NIR. This is seen clearly in Fig~\ref{Asymmetry}, which shows a comparison in both H- and J-bands between the SPHERE data (top row), and the raytraced images of M0 (second row). The figure also shows the raytraced images of models M1 and M2, to be discussed below in Sections \ref{alpha} to \ref{misallignment} (third and bottom rows, respectively). Reproducing the structure seen in the SPHERE/IRDIS images requires the localized modification of some parameters to account for three major differences between these images and the raytraced images of model M0 at NIR wavelengths. In first place, the H- and J-band images of M0 show clearly the presence of the second ALMA ring, which is not visible in our NIR data. Secondly, the produced images of model M0 show a more extended RID than observed in the data. And lastly, since model M0 is azimuthally symmetric, it fails to reproduce the observed asymmetry along the major axis of the first ring, described in Sect. \ref{Section3}. The modification of parameters required to reproduce the SPHERE/IRDIS images, and the two models resulting from these modifications, will be discussed in Sect. \ref{alpha} through \ref{misallignment}.

\subsection{Absence of outer rings in polarized scattered light: increased settling or grain growth in the outer disk?}\label{alpha}

Compared to the ALMA image, the polarized scattered light images in both the H and J band show a curious absence of emission beyond the ring at $\sim$0.63 arcsec. While the rings seen in the ALMA image are bright and emission is measured out to almost three times the radial extent of the scattered light disk, no structure is seen in polarized scattered light in the outer disk with an SNR above $\sim$2.5. Considering scattered light has also been detected out to 500 AU with roll subtraction \citep{2003SPIE.4860....1S, 2004PASP..116...55F} by \citet{2000ApJ...544..895G} using \textit{HST} STIS and by \citet{2008ApJ...682..548W} with \textit{HST} ACS, we are left to wonder as to the reason for the absence of polarized intensity observed in our SPHERE images beyond the first ring. While the sensitivity of our ground-based observations is lower than that of \textit{HST}, model M0 shows that after modelling the surface density of the outer disk the second ring should still be clearly visible at 1.25 and 1.6 $\mu$m even considering only $\sim 30\%$ of scattered light is polarized by dust. Considering the properties detailed for our model in Sect. \ref{model}, and assuming no abrupt radial variations in these parameters, a sharp decrease in surface density would be required at the location of the second gap, around $\sim$0.85 arcsec or 105 AU, in order to obscure the outer rings. However this is not observed in the results obtained from the modelling of the surface density based on the ALMA continuum image.

We can infer from this that some of the input parameters for our model are in fact not radially constant. Since we know there is a large dust mass at radii larger than 100 AU, it follows that there are no, or few, small grains present in the surface of the disk directly illuminated by the star in this region. This can be explained by the small grains settling closer to the midplane of the disk, in the shadow cast by the first ring, or by a local depletion of the smallest grains starting at around 100$\pm$15 AU from the center of the disk. 

Our next model (M1) attempts to explain the absence of outer rings in the J and H-band images using a modified grain size distribution in Zone 2c. In order to hide the outer rings in this zone while maintaining turbulence constant throughout the disk the optical depth needs to be decreased. We do this by depleting this region of the smallest dust grains, which are responsible for the optical depth in the NIR. We implement this in our first model by increasing the minimum particle size while keeping $a_{\rm max}$ and $a_{\rm pow}$ unchanged. With this modified setup we then execute the iterative procedure described in Sect. \ref{surfdensmod}. We find that depleting this zone of grains up to a size of 3 $\mu$m is necessary to decrease the optical depth in this region enough to obscure the second and third rings in the H-band and J-band Q$_{\phi}$ images. We find a depletion of all small grains up to 2.5 $\mu$m still allows for the second ring to show up clearly in the raytraced images of our models. In Sect. \ref{2poppy} we present a dust evolution model that shows it is possible to obtain a rapid decrease in the surface density of small dust grains, and we show this partial depletion is strong enough to almost completely obscure the presence of a second ring at 125\,AU.

\begin{figure}
  \centering
    \includegraphics[width=88mm,left]{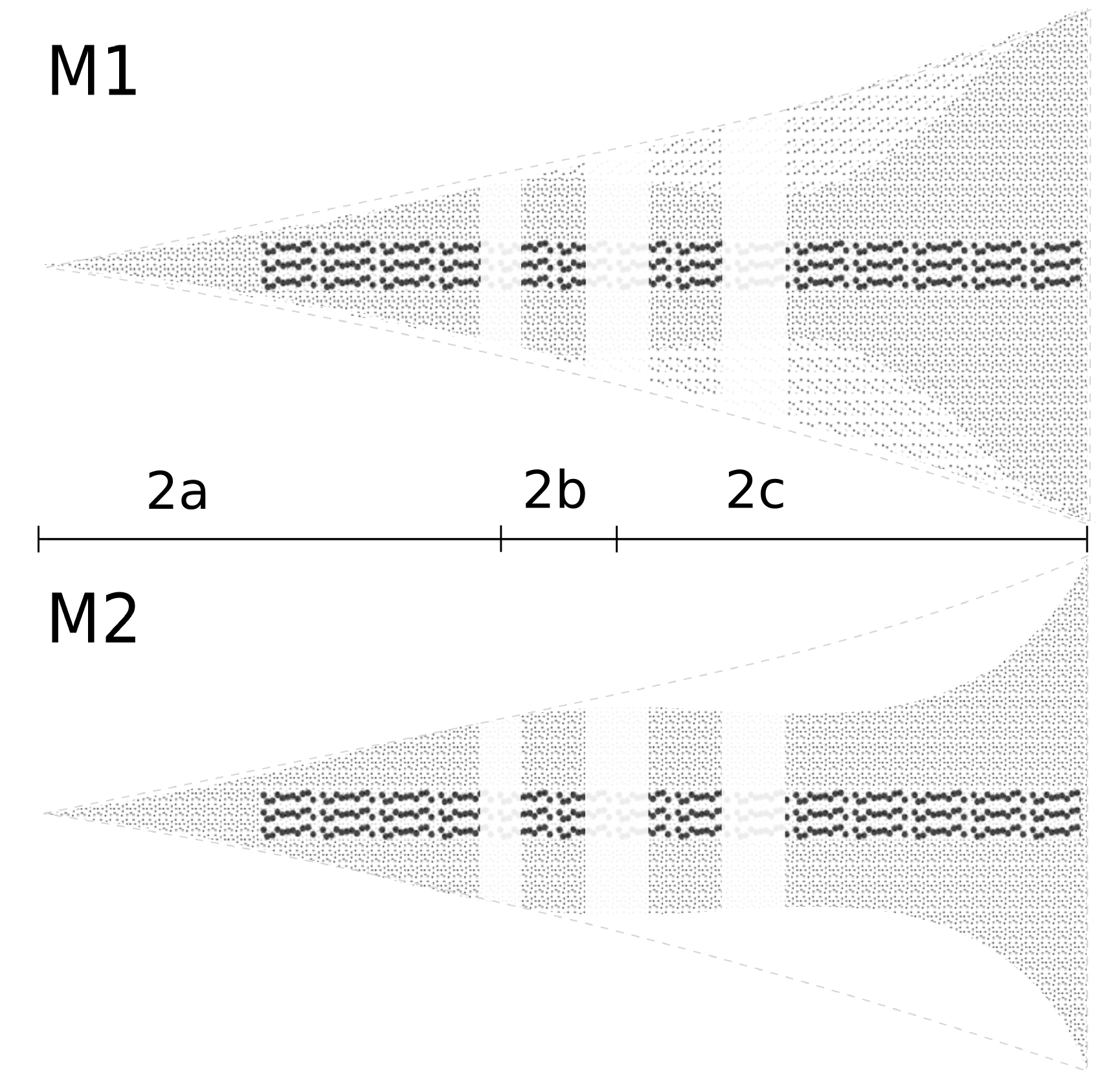}
  \caption{Simplified graphical representation (not to scale) of the two possible scenarios that might explain the lack of observed polarized scattered light in the SPHERE/IRDIS images in the outer disk. Top: small grain depletion scenario, in which small grains are present throughout the inner regions of a flared disk but are largely depleted on the surface of the disk at the location of the second ring. The larger grains are shown on the disk midplane, while the locations of the gaps are indicated by the white vertical bands. Bottom: increased settling scenario showing a double flaring structure for the dust and therefore for the scattering surface. A flaring outer edge with a small number of small grains remaining is necessary in both scenarios to account for the scattered light observed in \textit{HST} images.}\label{geom2}
\end{figure}

\begin{figure}
 \centering
  \includegraphics[width=88mm,left]{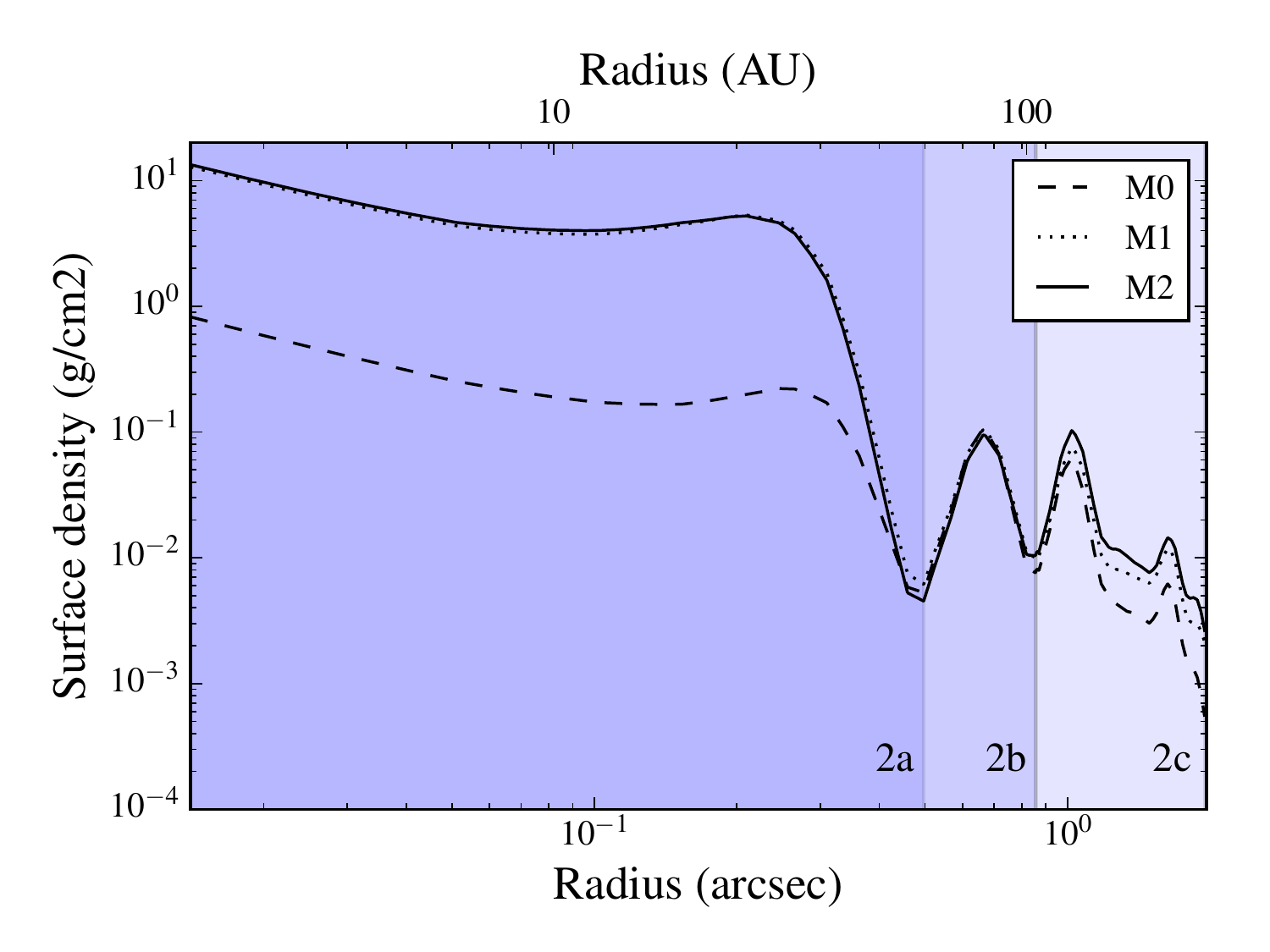}
  \caption{Surface density profiles for models M0 (dashed), M1 (dotted) and M2 (solid). Zones 2a, 2b and 2c are indicated in the figure as progressively lighter shades for the background 
  color.}\label{surfdens}
\end{figure}

Our last model (M2) uses increased settling of the dust grains at radii larger than 105 AU. This is achieved by decreasing the turbulence in Zone 2c to bring the small grains on the surface of the disk closer to the midplane and into the shadow cast by the first ring (zone 2b), while keeping the grain size distribution in this zone equal to the rest of Zone 2. With this modified setup we then execute the iterative procedure described in Sect. \ref{surfdensmod}, just as we did for model M1. We find that the $\alpha$ parameter needs to be lowered to $\sim1\times10^{-5}$ in order to achieve this effect. While we model this as a sharp decrease in the turbulence at 105 AU for simplicity, a more gradual decrease across the gap separating zones 2b and 2c would yield similar results. The global distribution of the small dust grains in these two models is summarized in Fig~\ref{geom2}. The top panel shows a flaring disk with a localized depletion of small grains in a ring around the disk, starting in our case somewhere in between zones 2b and 2c, and extending out indefinitely, possibly all the way to the outer edge of the disk, corresponding to the geometry of model M1. The bottom panel shows the geometry of the disk for model M2, consisting of a flared structure that extends out to the first ring at 77 AU, a decrease in the height of the scattering surface caused by localized increased settling in zone 2c. While both models are on their own sufficient to explain the absence of scattered light observed in the HD\,163296 disk beyond $\sim$100 AU, they are not completely independent. Even a partial depletion of small grains in the outer disk would cause the height of the $\tau=$ 1 surface to decrease in some degree, resulting in a geometry that is effectively very similar to that of the first model, albeit with a different mechanism at work behind it. Additionally, increased settling and depletion of small grains in this zone are not mutually exclusive, and the absence of the outer dust rings in the scattered light images might well be caused by a combination of these two effects, or in combination with a decreased pressure scale height or dust chemistry. The obtained surface density profiles for these models are shown in Fig~\ref{surfdens}, along with the total mass of zone 2c obtained for each model in Table \ref{Tab2}. Here we can see the effect of modifying the grain size distribution and turbulence in zone 2c: by increasing the minimum particle size in the outer disk (M1), we are effectively increasing the mass of this component of the disk by $\sim50\%$ during the modelling of the surface density either as a product of a decreased temperature or optical depth. Since the smallest dust grains provide only a small fraction of the optical depth at millimeter wavelengths (a few percent), we can attribute this to a decreased temperature caused by the diminished absorption of stellar light in this zone. Similarly, the mass in this zone is approximately doubled if instead the turbulence is decreased to $\alpha=1\times10^{-5}$, also as a product of a lowered midplane temperature (M2). 

\subsection{The resolved inner disk}\label{RID}

One of the most evident differences between the observations at millimeter and NIR wavelengths is the apparent size of the RID (zone 2a). The RID appears about 50$\%$ larger in the ALMA image than in the H-band polarized scattered light image, and even larger compared to the J-band image (see Fig~\ref{Asymmetry}).

In order to reduce the apparent size of the RID in models M1 and M2, the scale height of zone 2a was modified. The scale height was fixed at the inner radius of zone 2a, but the flaring exponent p was reduced from p= 1.15 to p= 1.07, making this zone close to wedge-shaped. Decreasing the scale height of this zone in this manner has the desired effect of reducing the apparent size of the RID, but it also modifies its temperature structure. By decreasing the scale height we are also reducing the amount of stellar light being absorbed by the dust in this component of the disk, which makes the temperature in this region drop. The drop in temperature translates directly to a drop in the surface brightness, and therefore to a larger dust mass obtained in this zone as a result of the iterative modelling of the surface density described in Sect. \ref{surfdensmod}. The surface density in this zone is found to increase by over an order of magnitude in models M1 and M2 as a product of the reduced scale height (see Fig~\ref{surfdens} which shows the surface density profiles in zone 2 for models M0, M1 and M2), yielding a total dust mass for the whole disk of $\sim 2.6 \times 10^{-3}$ M$_{\odot}$, which seems unusually high. Accurately estimating the mass of zone 2a is a complicated matter, considering this zone is optically thick in our models. However, due to the large dust masses required in this zone for models M1 and M2, it seems unlikely that a reduced scale height of this zone is the sole responsible for its smaller apparent size in the scattered light images.  

The total mass of the disk, and of its individual zones, are detailed for each of the three models in Table \ref{Tab2}. Interestingly, the surface density in zone 2b remains approximately equal in all three models, showing that the temperature structure on this zone remains independent of the modified scale height in zone 2a.

\begin{table}
 \caption{Total mass for each zone for all three models, in solar masses, after the iterative modelling of the surface density.}\label{Tab2}
 \centering
 \begin{tabular}{lccc}\hline
     & M0 & M1 & M2\\\hline\hline 
 Zone 1 & $6.44\times 10^{-9}$ & $6.44\times 10^{-9}$ & $6.44\times 10^{-9}$\\\hline
 Zone 2a & $1.3\times 10^{-4}$ & $2.3\times 10^{-3}$ & $2.25\times 10^{-3}$\\\hline
 Zone 2b & $1\times 10^{-4}$ & $1.1\times 10^{-4}$ & $9.8\times 10^{-5}$\\\hline
 Zone 2c & $1.4\times 10^{-4}$ & $2.1\times 10^{-4}$ & $2.8\times 10^{-4}$\\\hline
 Total & $3.7\times 10^{-4}$ & $2.6\times 10^{-3}$ & $2.6\times 10^{-3}$\\\hline
 \end{tabular}
\end{table}

\subsection{Brightness asymmetry: evidence of self-shadowing?}\label{misallignment}

Another predominant feature of the scattered light images is the asymmetry observed in the flux density of the ring (zone 2b) along the major axis, which cannot be explained by either the scattering phase function or the angle dependence of the polarization efficiency. Such an asymmetry could be caused by an accumulation of dust on the North side of the disk, or by the shadow of an inclined inner disk on the South side of the ring, similar to what has been proposed for HD\,142527 \citep{2015ApJ...798L..44M}, HD\,135344B \citep{2016A&A...595A.113S} and HD\,100453 \citep{2017A&A...597A..42B}. Considering this asymmetry is not observed in the ALMA image, we favor the shadowing scenario. 

Taking into account the structure of HD\,163296 two possibilities are available, with either the zone 1 (UID) inclined by a few degrees with respect to the rest of the disk, or with both zones 1 (UID) and 2a (RID) misaligned by similar degree. We find that an inclination of 3$^{\circ}$ with respect to the rest of the disk for zone 1, characterized by an inclination of 45$^{\circ}$ and a position angle of 136$^{\circ}$, produces a brightness asymmetry qualitatively very similar to the one observed in the scattered light images. This inclination however also has the effect of casting a shadow on the south side of zone 2a (RID). If both zones 1 and 2a of the disk are slightly inclined, however, this shadow is avoided. We find that if both components are inclined, an inclination relative to the outer disk of just $\sim$1.3$^{\circ}$, characterized by an inclination of 45$^{\circ}$ and a PA of 133.2$^{\circ}$, is required to reproduce to a high degree the asymmetry seen in the scattered light images. 

Fig~\ref{Asymmetry} shows the H- and J-band Q$_{\phi}$ images of the SPHERE data (top) and the obtained raytraced Q$_{\phi}$ images of models M0, M1 and M2 (second, third, and bottom row, respectively). These images were convolved with Gaussian kernels with a FWHM of 41 and 50 milliarcseconds respectively to simulate the resolution of the J and H-band images, measured from the PSF of the flux images. The models were scaled down such that the integrated flux density in an elliptical annulus containing the ring, with a width of 4 times the image resolution, equals that of the observations. The noise was measured from the H-band and J-band U$_{\phi}$ images in apertures with a radius of 4 and 3.4 pixels, respectively, assuming they contain no remaining signal, and added artificially to the model images pixel-by-pixel using a Gaussian random number generator to simulate the noise. In model M1 zone 1 has been inclined with respect to the rest of the disk to show the produced asymmetry. In model M2 both zones 1 and 2a have been inclined as described to show the resulting asymmetry. The choice of inclining different zones in these two models was motivated by purely illustrative purposes, and has no larger significance. The purpose is to show that similar results can be obtained by inclining only zone 1, or both zones 1 and 2a. Either choice of inclined components would work for either model.

\begin{figure}
  \centering
    \includegraphics[width=88mm,left]{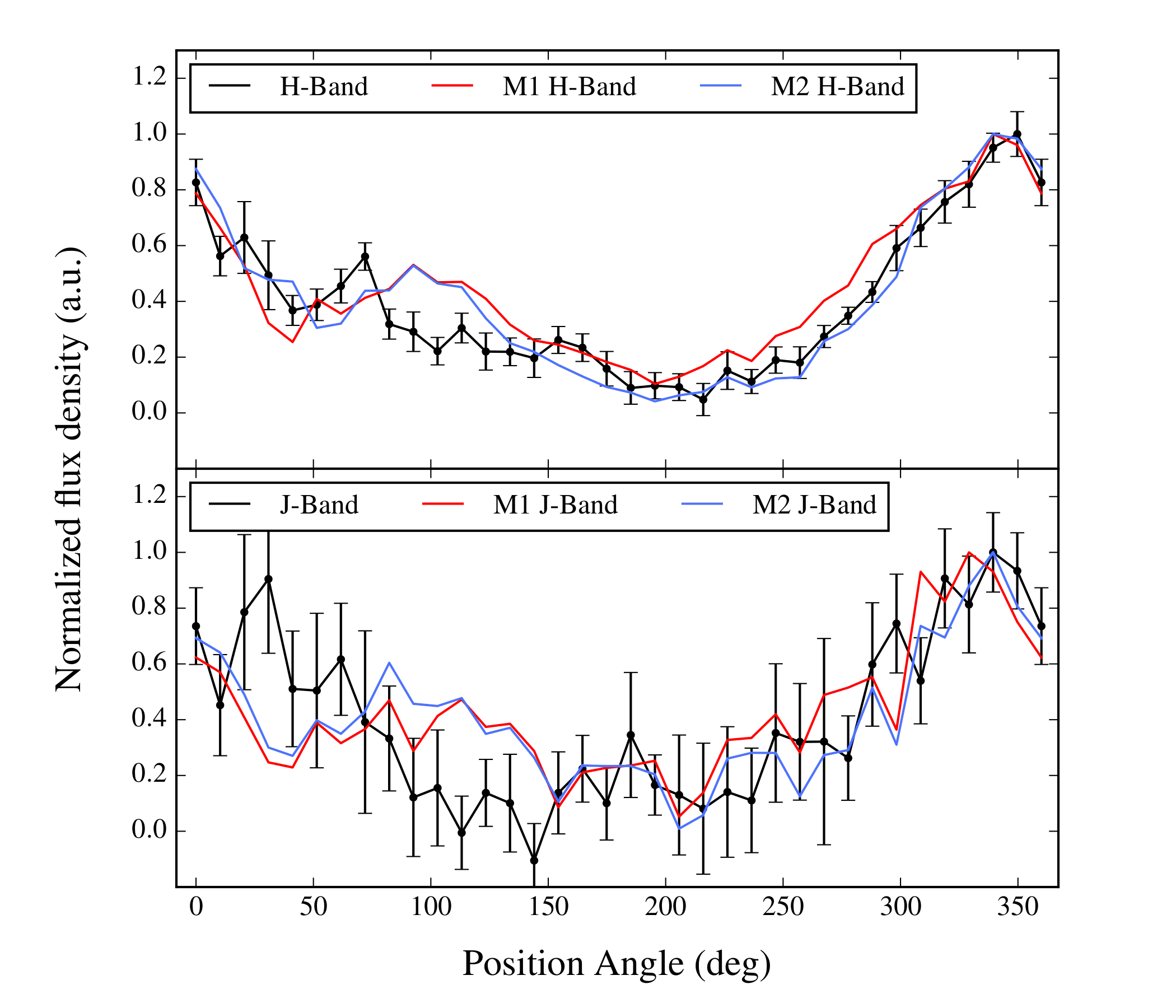}
  \caption{Azimuthal profiles of the ring for the SPHERE data (black), model M1 (blue) and model M2 (green) for both H-band (top) and J-band (bottom). The profiles were measured at a distance of 77 AU and 74 AU for H- and J-band images, respectively.}\label{profile_models}
\end{figure}

\begin{table}[b]
 \caption{Parameters modified in models M1 and M2 compared to the parameters used in model M0.}\label{Tab3}
 \centering
 \begin{tabular}{lccc}\hline
   & M0 & M1 & M2\\\hline\hline 
 $\alpha$ in zone 2c 	& $2.4\times 10^{-3}$ 	& $2.4\times 10^{-3}$ 	&  $1\times 10^{-5}$ \\\hline
 a$_{\rm min}$ ($\mu$m) in zone 2c & $5\times 10^{-2}$ & 3 & $5\times 10^{-2}$   	\\\hline
 H$_{0}$ (AU) in zone 2a & 5.64 & 0.073 & 0.073 	    				\\\hline
 r$_{0}$ (AU) in zone 2a & 100 & 1.7 & 1.7\\\hline
 p in zone 2a & 1.15 & 1.07 & 1.07 \\\hline
 $i$ zone 1 & $46^{\circ}$ & $45^{\circ}$ & $45^{\circ}$\\\hline
 PA zone 1 & $132^{\circ}$ & $136^{\circ}$ & $133.2^{\circ}$\\\hline
 $i$ zone 2a & $46^{\circ}$ & $46^{\circ}$ & $46^{\circ}$ \\\hline
 PA zone 2a & $132^{\circ}$ & $132^{\circ}$ & $133.2^{\circ}$\\\hline
 \end{tabular}
\end{table}

The resulting azimuthal profiles of models M1 and M2 are shown for both H- and J-band in Fig~\ref{profile_models} along with the SPHERE profiles for comparison. Both models show a similarly-good agreement with the J-band profile in the third and fourth quadrant of the disk, with some deviations in the first and second quadrant. However, the SNR in this band is much lower than in the H-band and therefore should be interpreted with care. For the H-band, on the other hand, we can see an excellent agreement with the data for both models, which manage to reproduce the observed asymmetry to a high degree except at postion angles between $\sim50^{\circ}$ and $130^{\circ}$. The reason for this difference is currently unclear.  It might be due to a local disturbance in the scaleheight of the disk, either in the ring, or in the inner regions causing a localized shadow on the ring.  It is also possible that the dust properties or the abundance of small grains are different at this location, but a structure like this would be expected to wash out within a few orbital time scales.

\begin{figure}[t]
  \centering
    \includegraphics[width=88mm,left]{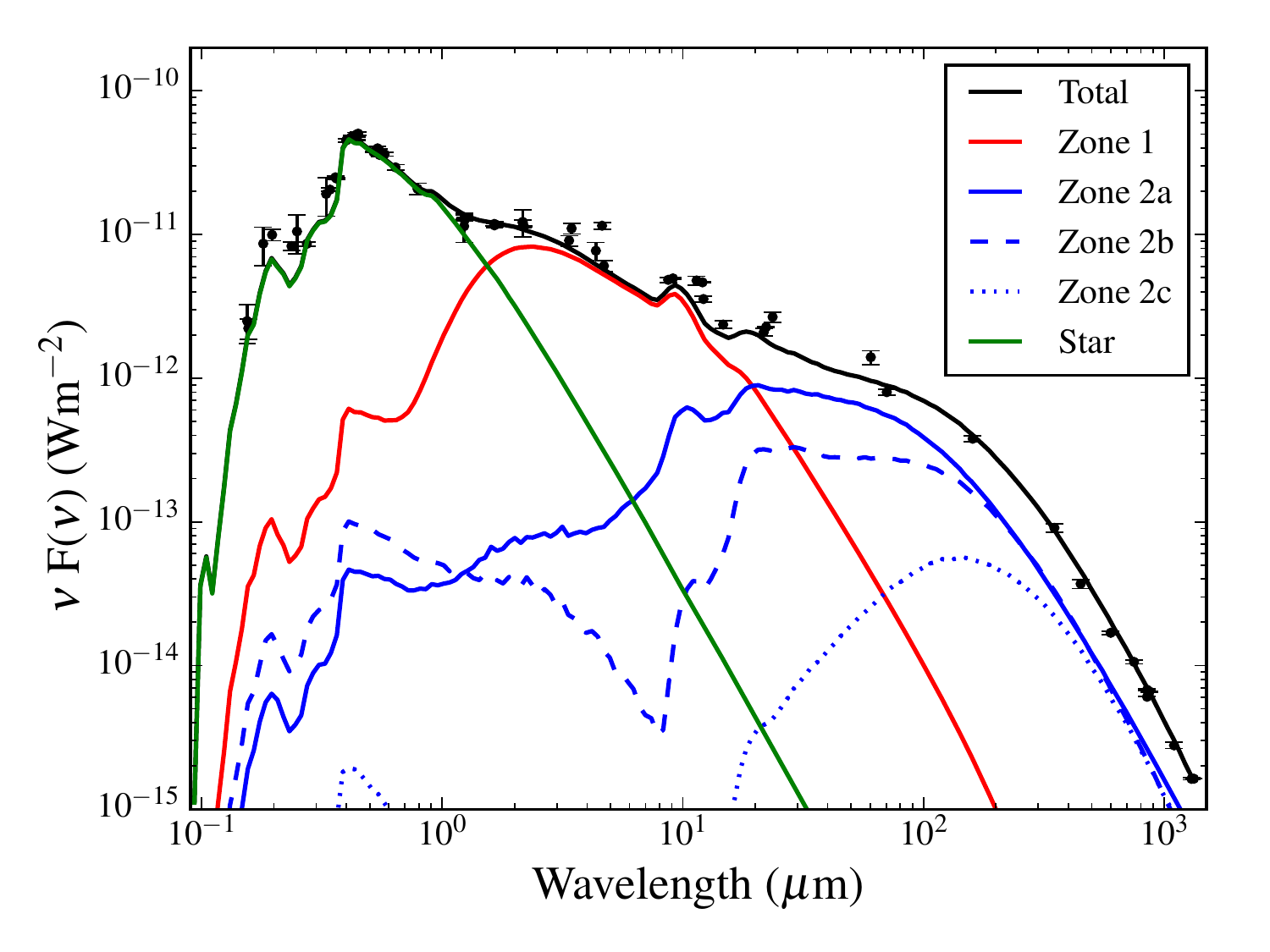}
  \caption{Spectral energy distribution of HD\,163296. The points and their corresponding error bars correspond to the photometry of the source. The total SED of model M2 is shown in black, along with its individual components: stellar spectrum (red), zone 1 (blue), zone 2a (green, solid), zone 2b (green, dashed) and zone 2c (green, dotted).}\label{SED}
\end{figure}

A summary of the parameters used for each model is found in Table \ref{Tab3}.

\section{Discussion}\label{Section5}

\subsection{Disk geometry}

While typically classified as a group II source, HD\,163296 is often referred to in the literature (e.g. \citealt{2003A&A...398..607D}) as an intermediate type source. It is tempting therefore to assume it might be an example of a type of protoplanetary disk that constitutes an evolutionary link between typical group I and group II sources. While this is indeed a possibility the true nature of the evolution of Herbig disks might be different. 

Two separate evolutionary processes should be distinguished in non-compact disks: the formation of an extended gap or cavity in the outer disk leading to group I-like SEDs, and the deflaring or vertical collapse of the disk, leading to flat group II sources, assuming all disks start out as both continuous and flaring. It follows from observational evidence that the disk around HD\,163296 is neither compact nor flat or -- at least not completely -- self-shadowed, as it can be observed in scattered light images as far out as $\sim$500 AU. On the other hand observations of the source so far have failed to resolve a cavity in the outer disk, either in millimeter dust emission or scattered light (though cavities have been detected in $^{13}$CO and C$^{18}$O with ALMA out to a few tens of AU \citep{2016PhRvL.117y1101I}, comparable to the cavity sizes of group I sources). We may tentatively conclude then, until further observational evidence, that HD\,163296 constitues part of a sub-category of protoplanetary disks that has undergone neither of the processes that characterize either group I or group II sources. It follows that the disk either constitutes part of a separate group of protoplanetary disks, or that it's in an evolutionary stage anterior to either group I or group II sources.

\begin{figure}[ht]
  \centering
    \includegraphics[width=88mm,left]{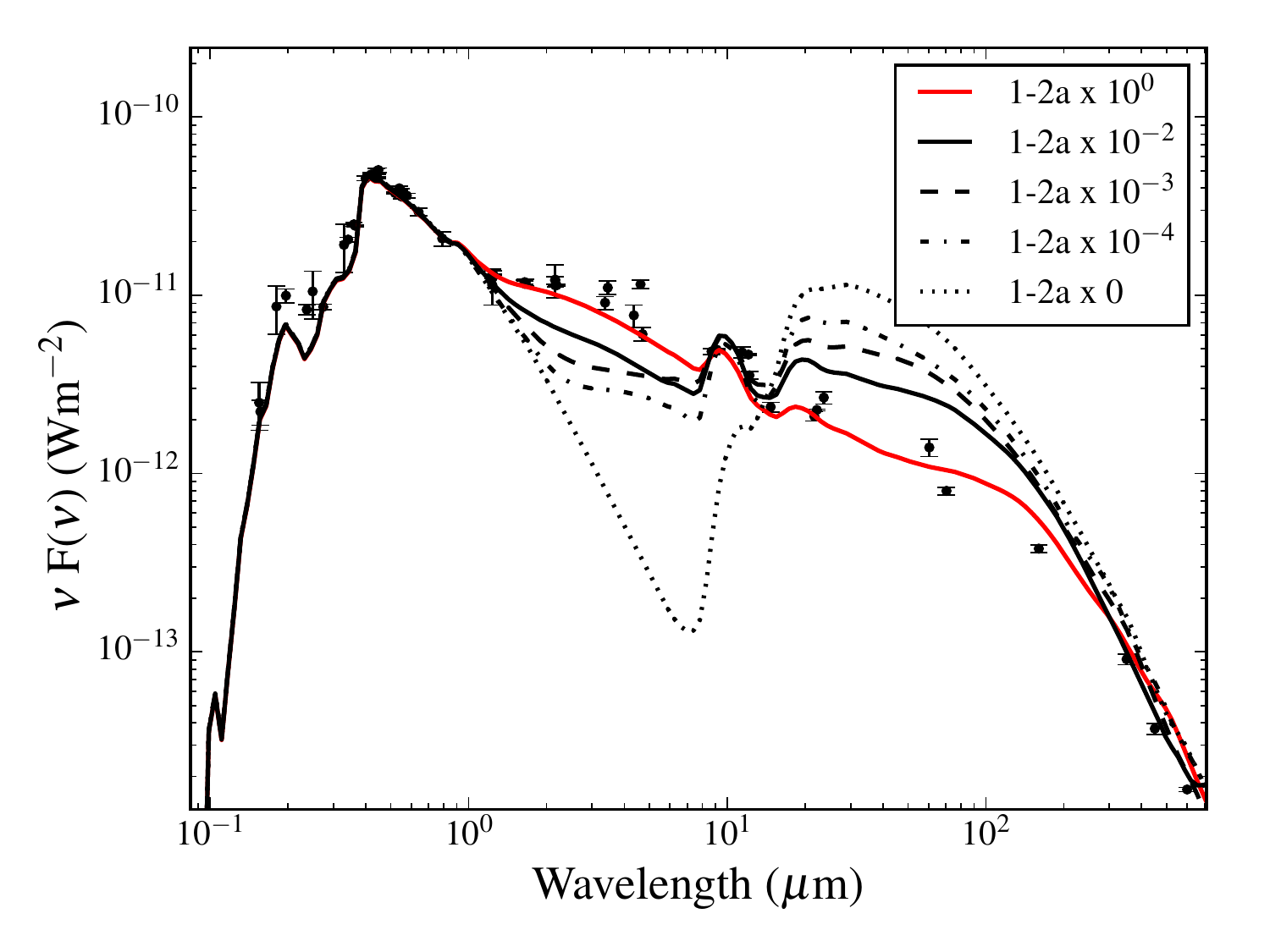}
  \caption{Spectral energy distribution of HD\,163296. The points and their corresponding error bars correspond to the photometry of the source. The total SED of model M2 is shown as a thick solid red line. The solid, dashed, dot-dashed and dotted black lines show the SEDs obtained by depleting zones 1 and 2a of the model by a factor $10^{2}$, $10^{3}$, $10^{4}$, and complete depletion, respectively.}\label{SED2}
\end{figure}

With this in mind we computed the SED for each of the components of model M2, which can be seen in Fig~\ref{SED}, by raytracing our model at 200 different wavelengths between 0.1\,$\mu$m and 7.5\,mm. The SED obtained for model M1 is qualitatively very similar. The figure shows the photometry of HD\,163296, along with the total SED of the model and the SED contribution of each of its individual components. The SEDs of models M0 and M1 are very similar, since the initial model used fits the SED and the modifications implemented in the models do not have a big effect on it. We can see the bulk of the NIR flux comes from the thermal emission of zone 1, with only a small fraction attributed to zones 2a and 2b in scattered light. In the FIR the thermal emission from zone 2a dominates, with a weaker but still comparable contribution from the thermal emission of zone 2b, smaller only by about a factor 2 up to $\sim$100 $\mu$m, and about equal from there up to millimeter wavelengths. This seems to suggest zones 2a and 2b combined might be playing a role in the spectral energy distribution of HD\,163296 similar to that of the bright inner wall of the outer disk in group I sources.

To analyze this further, we generated four different models identical to M2 in all but the total dust mass in zones 1 and 2a (UID and RID). The dust in these zones of the first, second and third of these models was depleted by a factor $1\times10^{2}$, $1\times10^{3}$, and $1\times10^{4}$, respectively. In the fourth model the dust in zones 1 and 2a was depleted completely. The resulting SED for all these models is shown in Fig~\ref{SED2}. The figure illustrates the effect in the spectral energy distribution of progressively removing the inner disk components, effectively creating a large cavity in the disk inside the first ring. Not only does the NIR component of the total disk SEDs decrease as dust is removed from both the inner disks, but the thermal emission of the outer disk increases as the temperature in this zone rises, a consequence of progressively increasing the amount of stellar radiation received by this zone. Lost NIR excess notwithstanding, our models show that artificially creating a dust cavity in the disk, even a partially-depleted one, can produce a larger FIR excess reminiscent of typical group I sources. This leads us to believe that the lack of such a cavity might be the largest discriminator between HD\,163296 and other group I sources.

\subsection{Absence of multiple rings in scattered light}\label{2poppy}

\begin{figure}[ht]
  \centering
    \includegraphics[width=88mm,left]{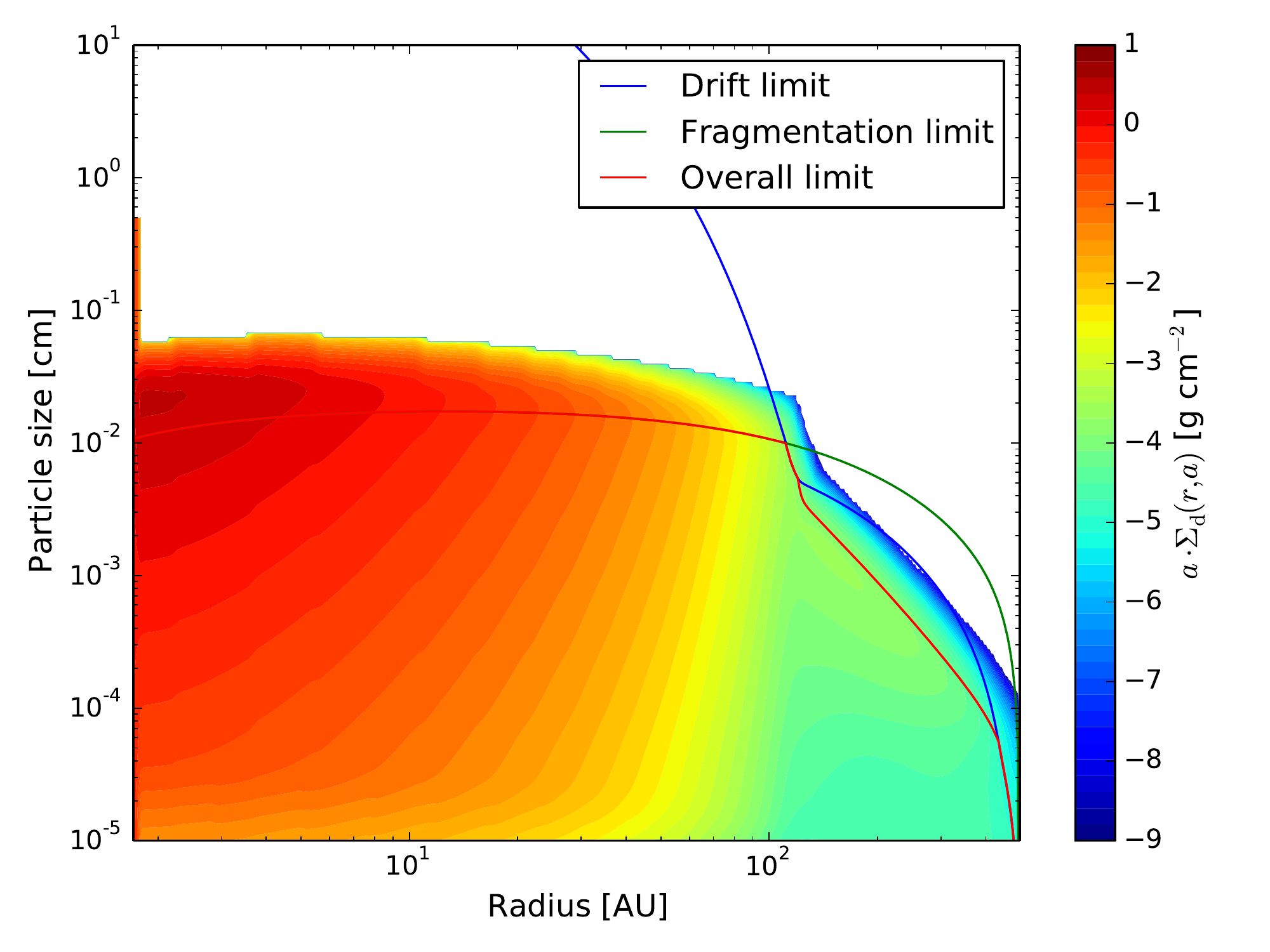}
  \caption{Reconstructed grain size distribution for a 2.5 Myr disk with an initial mass of 0.2 M$_{\odot}$, initial dust-to-gas ratio of 0.02, turbulence $\alpha=2.4\times10^{-3}$ and fragmentation velocity $v_{frag}=$ 3 m/s. }\label{twopoppy}
\end{figure}

\begin{figure*}[t]
  \centering
    \includegraphics[width=180mm,left]{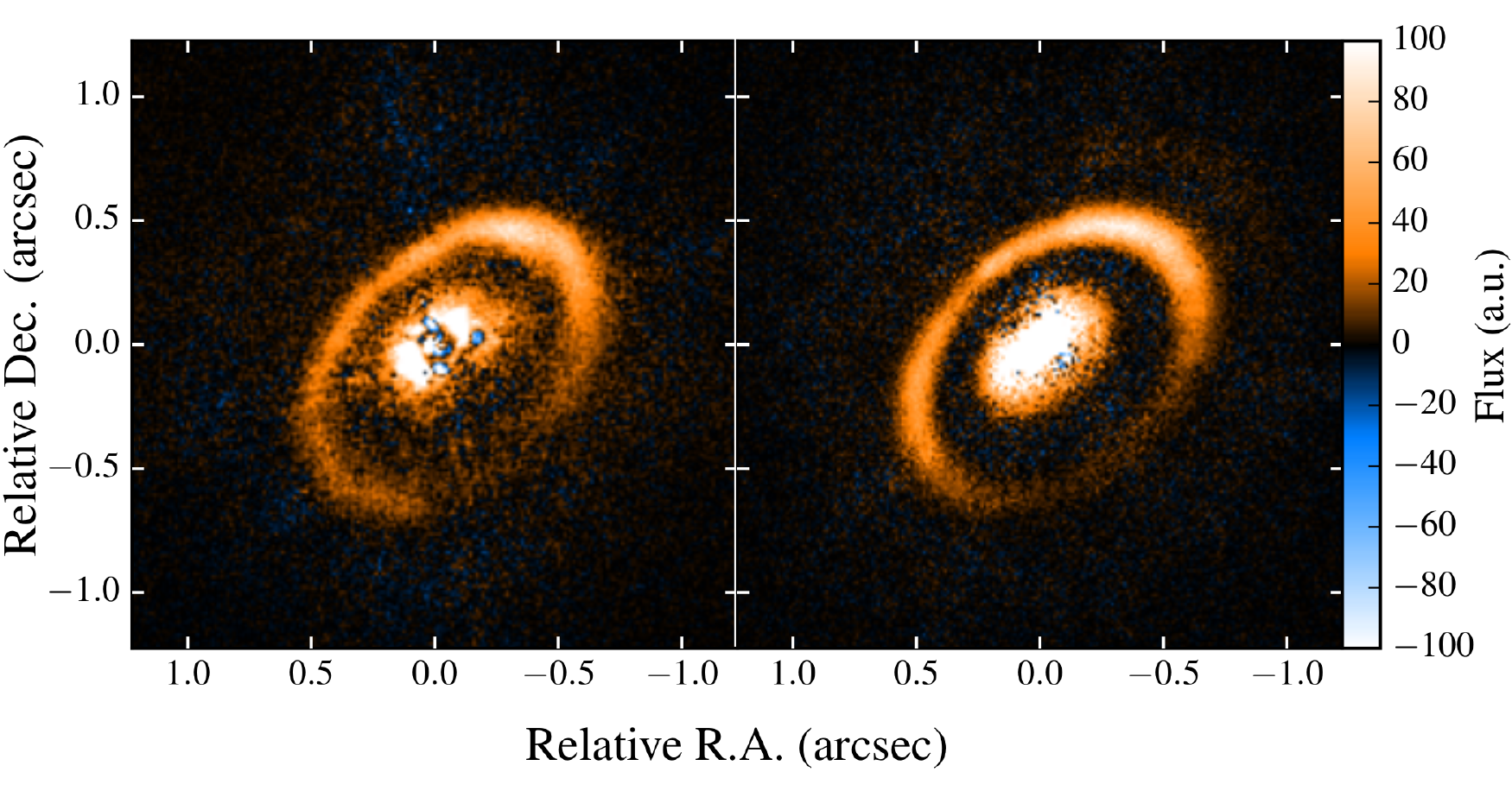}
  \caption{Comparison between the H-band Q$_{\phi}$ SPHERE image (left) and model M2 (right). A second ring was artificially added to the image of the model with a semi-major axis of 125\,AU, a thickness of 24\,AU, and a surface brightness 46 times lower than that of the first ring. }\label{ring2}
\end{figure*}

Three bright rings at $\sim$75, 125 and 200\, AU from the center of the disk can be seen in the thermal emission from the midplane, while our polarized scattered light images with IRDIS in the J and H-bands show a clear lack of the two outermost of these rings. Two models were presented in Sect. \ref{alpha} capable of reproducing the observations, explaining the lack of small dust grains on the disk surface beyond $\sim100$\,AU by either an increased settling (model M2) or a depletion of small grains (model M1) in the outer disk (zone 2c). While we modelled this depletion by completely removing all dust grains smaller than $3\, \mu$m in zone 2c, we aim to show that a strong but partial depletion of the smallest dust grains in the outer disk is not only capable of reproducing the absence of the outer rings in scattered light, but is also consistent with dust evolution models. 

As a proof of concept we present the results from a dust evolution simulation obtained using the dust evolution code \textit{twopoppy} \citep{2012A&A...539A.148B}. Our goal was not to produce an accurate model of the surface density of the disk, but rather to show there may be regions in a disk where a depletion of small grains can be achieved naturally as a product of dust evolution. The code was used to evolve a disk with an initial mass of 0.2 M$_{\odot}$. A turbulence parameter of $2.4\times10^{-3}$ was used for consistency with our \textit{MCMax3D} model, along with a low fragmentation velocity of 3 m/s and an initial dust to gas ratio of 0.02. An initial gas surface density characterized by the parameter $\gamma=$ 0.8 and a characteristic radius r$_{c}=$ 165 AU was used, with:

\begin{ceqn}\label{sigma3}
 \begin{align}
    \Sigma(r) \propto r^{-\gamma} \exp \left( - \left( \frac{r}{r_{c}} \right)^{2-\gamma} \right) .
 \end{align}
\end{ceqn}

A high initial disk mass and dust-to-gas ratio are required to obtain a total dust mass of $5\times10^{-4}$ M$_{\odot}$, in the lower end of the $5-17\times10^{-4}$ M$_{\odot}$ dust mass range found in the literature \citep{2004A&A...416..179N, 2008ApJ...689..513T, 1997ApJ...490..792M, 2007A&A...469..213I}, and a total disk mass of $\sim7\times10^{-2}\,M_{\odot}$ after 2.5 Myr of evolution. Due to the rate at which the disk loses dust mass it is not possible to achieve a dust mass of $5\times10^{-4}$ M$_{\odot}$ after 5 Myr of evolution without a much higher initial disk mass or, alternatively, a much higher initial dust-to-gas ratio. A way of achieving the larger dust mass observed in the disk around HD\,163296 without requiring a very high initial dust-to-gas ratio or initial mass is with a disk that has continued accreting gas and dust from the circumstellar environment for a couple million years after its formation, replenishing the dust content of the disk and allowing this way for the large dust mass we observe today. If this were the case for HD\,163296, this would mean the disk we are observing is effectively younger than the 5 Myr estimated for the system. Evolving such a disk for 2.5 Myr produces the grain size distribution seen in Fig~\ref{twopoppy}, obtained using the grain size distribution reconstruction module of \textit{twopoppy} \citep{2015ApJ...813L..14B}.

The figure shows a grain size distribution that is limited by fragmentation in the inner $\sim$100 AU of the disk, and mostly by grain growth in the outer disk, except for a small transition zone in between them which is radial drift-limited. The small grains in the inner region grow quickly to the fragmentation limit, and are then replenished by the collision and fragmentation of the larger grains. In the outer disk dust grains have grown quickly to the drift limit and drifted inwards, leaving behind a population of small grains with a density too low to grow efficiently to larger sizes. This means the smaller grains in this region can still grow slowly, but are not efficiently replenished as they don't reach the fragmentation limit. In the intermediate transition region, on the other hand, dust is still able to grow to millimeter sizes, but the drift limit is reached before they can fragment, making the replenishing of small grains very inefficient. 

Since the outer disk is growth-limited, we observe no large amount of large grains beyond 100-200 AU, qualitatively consistent with the ALMA observations, which show a sharp decrease in optical depth around $\sim$125 AU between the second and third rings. The remaining intermediate size grains beyond $\sim$125\,AU could also provide the low optical depth responsible for the faint thermal emission seen from the third ring in the ALMA image. The low abundance of small grains remaining in the outer disk in this model might be enough to explain the emission seen in the \textit{HST} scattered light images, while at the same time being low enough to explain the absence of the second and third ALMA rings in the polarized scattered light images. 

To determine if this depletion of small grains in the outer regions of the disk is enough to explain the non-detection of the second ring in the polarized scattered light images, we estimate the radial optical depth contrast between the location of the first ring, at 77 AU, and the second ring at 125 AU. The optical depth per unit length in the radial direction at a distance $r$ from the star in the J and H-bands is given by the sum of the optical depths contributed by each grain size population:

\begin{ceqn}
 \begin{align}
    \frac{\Delta \tau(r)_{J,H}}{\Delta l}  = \sum_{a_{i}} \kappa(a_{i})_{J,H} \rho(r,a_{i}) ,
 \end{align}
\end{ceqn}

\noindent with $\kappa(a_{i})_{H,J}$ the opacity of dust grains of size $a_{i}$ at the J or H-band, obtained from \textit{MCMax3D}, $\rho(r,a_{i})$ the density of grains of size $a_{i}$ on the surface of the disk at a distance $r$ from the star, and $\Delta l$ a radial distance. Since both rings are unresolved (or barely resolved) in the ALMA image, which has a resolution of 0.2 arcsec or $\sim$24\,AU, we cannot estimate their thickness. However, the FWHM for the first ring was measured from the H-band Q$_{\phi}$ image and found to be approximately 24\,AU. We shall assume for simplicity this is the thickness of both rings, though it is more likely an upper limit for the width of the second ring. This assumption will allow us to compare the optical depth contrast, and therefore the surface brightness contrast, between the two rings. We approximate $\rho(r,a_{i})$ by:

\begin{ceqn}
 \begin{align}
    \rho(r,a) \sim \frac{\Sigma(r,a)}{H(r)},
 \end{align}
\end{ceqn}

\noindent with $\Sigma(r,a)$ the surface density of particles of size $a_{i}$ at a distance $r$ from the center of the star, obtained from the grain size reconstruction of our dust evolution model, and $H(r)$ the \textit{MCMax3D} pressure scale height at a distance $r$ from the center of the disk. The optical depth contrast between the first and second ring can then be expressed as:

\begin{ceqn}
 \begin{align}
    \frac{\Delta \tau(R_{1})_{J,H}}{\Delta \tau(R_{2})_{J,H}} \sim \frac{H(R2)}{H(R1)} \frac{ \sum_{a_{i}} \kappa(a_{i})_{J,H}\Sigma(R1,a)}{ \sum_{a_{i}} \kappa(a_{i})_{J,H}\Sigma(R2,a) }, \label{tauuu}
 \end{align}
\end{ceqn}

\noindent with $R_{1}$ and $R_{2}$ the radii of the first and second rings, respectively. We integrated the product of $\kappa(a_{i}) \Sigma(r,a_{i})$ over all grain sizes at both radii, at a wavelength of 1.6 $\mu$m, and found this value to drop by a factor 14 from the first to the second ring. Accounting also for the spread of dust grains over a larger vertical extent (first term on the right-hand-side of Eq. \ref{tauuu}), using the scale height of our \textit{MCMax3D} model (see Table \ref{Tab1}), we find the optical depth at 125\,AU to drop by an additional 1.8 factor relative to the optical depth of the first ring, yielding a total optical depth contrast of about 25. 

Assuming the radial optical depth at the location of the first ring is $\tau \sim$1, this amounts to a relative intensity about 16 times lower for the second ring as a consequence of the drop in surface

\noindent density and the increasing scale height. Taking also into account the $r^{-2}$ dependence of the stellar radiation field, the total intensity of scattered light we could expect for the second ring in this model can be as low as 46 times fainter than the intensity of the first ring.

To test whether this contrast in intensity is high enough to account for the absence of the second ring in the scattered light images, we manually added a second ring to the H-band image of our model M2. This was done by measuring the average flux of the first ring at each position angle, scaling it down by a factor 46, and adding it to the image at a distance of 125\,AU from the center of the first ring to recreate a similar offset. The flux was measured at each position angle in circular apertures of radius $r=\,8$ pixels, and then added in circles of the same radius at the expected location of the second ring, and at the corresponding position angle. This in order to simulate a second ring with both a similar width to that of the first ring and a similar asymmetry. The resulting image is shown in the right side of Fig~\ref{ring2}. We see that, at $\sim2\%$ of the intensity of the first ring, the artificially-added second ring is almost completely invisible at the noise level of our polarized scattered light H-band image. Only the brightest side of the ring shows up, at a very low SNR, at the North-West side of the image. 

Even though this scenario can successfully explain the non-detection of the outer rings in the scattered light images, a depletion of the smallest grains in the outer disk to this degree is not strictly necessary to produce this effect. As discussed in Sect. \ref{alpha} an increased settling in zone 2c, characterized by a small turbulence $\alpha$ parameter, can also explain why the two outer rings observed in the ALMA image are not visible in polarized scattered light. With the small dust grains responsible for the opacity in the NIR settling closer to the midplane, the scattering surface can be pulled down and into the shadow cast by the inner ring. The radiative transfer models obtained using an $\alpha=\,1\times10^{-5}$ in Zone 2c (see Fig~\ref{Asymmetry}, bottom row) show that this increased settling can be successfully used to explain the absence of the outer rings in the scattered light images. 

While these two simplified cases are enough, on their own, to explain why we observe only the innermost ring in the scattered light images, a more complex model lying somewhere in between these two extremes is possible. A lower turbulence, coupled with a partial depletion of the smallest grains on the surface of the disk, can be expected to produce comparable results. An incomplete depletion of smaller grains in the outer disk is for us the favored scenario, since a small remaining number of small grains in the outer disk out to several hundred AU is required to explain the observations of scattered light out to $\sim$500 AU observed with \textit{HST}. Shadowing of the outer disk could also be caused by increasing the scale height in the region of the first ring. However, this would influence the SED and change the observed offset of the scattered light ring from the star.

\section{Conclusions}\label{Section6}

We have presented in this paper two independent models capable of reproducing simultaneously the millimeter continuum observations of HD\,163296's midplane with ALMA and polarized scattered light observations of the disk surface with SPHERE/IRDIS. 

The observations of the midplane has allowed us to obtain a surface density profile of the disk for the parameters defined in Sect. \ref{model}, out to a truncation radius of 240 AU. This truncation radius marks the outer edge of the dust disk at 1.3mm wavelength. The gas component of the disk, on the other hand, is seen to extend out to $\sim$500 AU in both gas line observations and scattered light. An inner disk component and a single ring at 77 AU characterize the polarized scattered light images in both J and H-band. The ring presents an offset of 105 mas, corresponding to a height of 12 AU above the disk midplane in the H-band and 11.5 AU in the J-band. This observation of scattered light is further evidence of a flaring disk. However, a simple flaring disk model is unable to account for the non detection of any signal beyond the second gap at ~100 AU in the SPHERE images.

The non-detection in scattered light of the two outer rings observed in the ALMA dust continuum image evidences the absence of small dust grains on the disk surface in this zone.  This points to either the smaller grains settling closer to the midplane, under the shadow cast by the first ring, or to a partial-to-complete depletion of the small grains in this region. We find with model M2 that a settling characterized by $\alpha \sim1\times10^{-5}$ is required to settle the small grains responsible for the opacity at J and H-bands into the shadow cast by the first ring. On the other hand, if settling is not increased with respect to the rest of the disk, a total depletion of grains smaller than 3 $\mu$m can have the same effect (model M1). The two mechanisms, while different, are approximately equivalent in that the produced effect is the lowering of the scattering surface under the surface of the disk directly illuminated by the central star. A dust evolution model has been provided as proof of concept that shows the feasibility of the depletion of small grains in the middle-disk, supporting the small grain depletion scenario.

The evidence of HD\,163296 possessing a global flaring geometry points to its misclassification as a group II source, insofar as our current understanding of these sources and their geometry goes. The lack of a largely-depleted inner cavity sets it apart from group I sources as well. With an intermediate type SED, we find from both of our models that the bulk of the FIR excess comes in similar measure from the resolved inner disk and the first ring. With the resolved and unresolved inner disks in the way, there is no large inner wall directly illuminated by the star, which likely explains why the FIR region of its SED is lower than a typical group I source. At the same time, the gap between the resolved inner disk and the first ring is reminiscent of the extended dust-depleted gaps or cavities commonly found in group I sources. Our models show depleting the inner regions of this disk does bring the properties of this object much closer to a group I source. Whether the disk is currently in the process of forming a dust-depleted inner cavity and transitioning towards a group I stage or not, both our models and observational evidence point to HD\,163296 having more in common with group I sources than with group II sources.  

The brightness asymmetry observed in the IRDIS images is likely the product of three separate effects: the scattering phase function, angle-dependent polarization efficiency, and self shadowing. We have provided a model that shows a small misallignment of $\sim$1.3$^{\circ}$ of the RID and UID suffices to reproduce this asymmetry to a high degree in all but the South-East quadrant, as well as a second model with the UID misaligned by 3$^{\circ}$ that produces similar results.

We may conclude from this work that multi-wavelength observations are fundamental for a thorough understanding and proper modelling of protoplanetary disks, due to the pronounced differences that arise at different wavelengths caused by the complex structures of disks. Some of the conclusions presented on this paper would not have been possible without the analysis of the disk observations at multiple wavelengths. 

\begin{acknowledgements}
SPHERE is an instrument designed and built by a consortium consisting of IPAG (Grenoble, France), MPIA (Heidelberg, Germany), LAM (Marseille, France), LESIA (Paris, France), Laboratoire Lagrange (Nice, France), INAF - Osservatorio di Padova (Italy), Observatoire de Genève (Switzerland), ETH Zurich (Switzerland), NOVA (Netherlands), ONERA (France), and ASTRON (The Netherlands) in collaboration with ESO. SPHERE was funded by ESO, with additional contributions from CNRS (France), MPIA (Germany), INAF (Italy), FINES (Switzerland), and NOVA (The Netherlands). SPHERE also received funding from the European Commission Sixth and Seventh Framework Programmes as part of the Optical Infrared Coordination Network for Astronomy (OPTICON) under grant number RII3-Ct-2004-001566 for FP6 (2004-2008), grant number 226604 for FP7 (2009-2012), and grant number 312430 for FP7 (2013-2016). G. M.-A. acknowledges funding from the Netherlands Organisation for Scientific Research (NWO) TOP-1 grant as part of the research programme ``Herbig Ae/Be stars, Rosetta stones for understanding the formation of planetary systems'', project number 614.001.552. M.B. acknowledges funding from ANR of France under contract number ANR-16-CE31-0013 (Planet Forming Disks).
\end{acknowledgements}

\bibliographystyle{aa} 
\bibliography{mylibrary} 

\end{document}